\documentclass[reprint,amsmath,amssymb,aps]{revtex4-1}

\usepackage{hyperref}
\usepackage{graphicx}
\usepackage{dcolumn}
\usepackage{bm}
\usepackage{colortbl}
\usepackage{url}

\begin{document}

\title{Spectral Photon Sorting For Large-Scale Cherenkov and Scintillation Detectors} 

\author{T. Kaptanoglu}
 \email{tannerk@hep.upenn.edu}
\author{M. Luo} 
\author{B. Land}
\author{A. Bacon}
\author{J. Klein}

\affiliation{University of Pennsylvania, 209 South 33rd Street Philadelphia PA 19104, USA}

\begin{abstract}

We describe here measurements with a new device, the ``dichroicon,'' a Winston-style light concentrator built out of dichroic reflectors,  which could allow large-scale neutrino detectors to sort photons by wavelength with small overall light loss. Photon sorting would benefit large-scale water or ice Cherenkov detectors such as Hyper-Kamiokande or IceCube by providing a measure of dispersion, which in turn could allow improved position reconstruction and timing. For scintillator detectors like JUNO, upgrades to SNO+ or KamLAND-ZEN, or to water-based liquid scintillator detectors like Theia, dichroicons would provide effective discrimination between Cherenkov and scintillation light, allowing them to operate as true hybrid detectors.

We include measurements with a prototype dichroicon using first a Cherenkov source to show spectral photon sorting works as expected.  We then present measurements of  two different LAB-based liquid scintillator sources, and demonstrate discrimination between Cherenkov and scintillation light. On the benchtop we can identify Cherenkov light with better than 90\% purity while maintaining a high collection efficiency for the scintillation light. First results from simulations of a large-scale detector are also presented.

\end{abstract}

\maketitle 

\section{Introduction}\label{sec:intro}

There is a rich history of discovery for large-scale neutrino detectors that use photons as their primary detection method~\cite{IMB,Ahn:2006zza,Fukuda:1998mi,Ahmad:2002jz,kamland,An:2012eh,Ahn:2012nd,Abe:2011sj,Adamson:2017gxd,Aartsen:2014gkd,Aguilar-Arevalo:2013pmq,Bellini:2011rx}.  These detectors are often monolithic, with a target medium in which Cherenkov or scintillation light is produced, viewed by an array of photon sensors.

Perhaps surprisingly, despite their great success to date, these detectors use only a small amount of the information available in the photons they detect.  A typical large-scale photon-based detector records at most the number of detected photons and their arrival times.  But photons may also carry information about physics events in their direction~\cite{Dalmasson:2017pow}, their polarization, and their wavelength.  

We focus here on the development of a device that is capable of providing information on photon wavelength in a large-scale detector.  In a Cherenkov detector---whether in water, ice, or oil---photon wavelength carries information about the propagation time from the source vertex to the photon sensor.  Across 50~m of water, for example, a 550~nm photon will arrive nearly 2~ns earlier than a 400~nm photon, easily resolvable by modern photomultiplier tubes (PMTs)~\cite{Kaptanoglu:2017jxo}. Thus, measuring the difference in time between many long-wavelength and short-wavelength photons that lie along a Cherenkov ring provides information about event position, independent from the overall timing and angular information usually used in reconstruction. The resolution of dispersion in such a detector also allows improved timing, as both the approximately 2~ns spread from dispersion and the differential effects of Rayleigh scattering broaden the prompt time window used for reconstruction.

In a scintillation or water-based scintillation detector, photon wavelength can be used to detect Cherenkov light independently from scintillation light. For these detectors, the scintillation light typically lies in a narrow band at short optical wavelengths, while Cherenkov light is naturally broadband. Future large-scale scintillation experiments like Theia \cite{Gann:2015fba} plan to detect both Cherenkov and scintillation light as a way of providing a very broad range of physics with a single detector. 

Independent detection of Cherenkov light allows reconstruction of event direction, which can aid in identifying solar neutrino events \cite{Bonventre:2018hyd,OrebiGann:2019ncf}, classification of neutrinoless beta decay candidates against the solar neutrino background \cite{Elagin:2016zgp,Jiang:2019cnb,Biller:2013wua}, or discrimination of high-energy $\nu_{e}$ events from $\pi_{0}$'s, which is important for studying long-baseline neutrino oscillations. The scintillation light provides a high light yield that is critical for good energy resolution and position reconstruction. The time profile of the scintillation light is also important, because it affects position reconstruction and provides ways of discriminating $\beta$s from $\alpha$ particles.

There are several possible techniques for measuring the Cherenkov light in liquid scintillator detectors. The timing of the detected photons is a powerful handle, as the Cherenkov light is produced promptly, whereas it may take as much as a nanosecond for the scintillation light to be emitted~\cite{Caravaca:2016ryf}. Additionally, the angular distribution of the Cherenkov light around the event direction distinguishes it from isotropic scintillation light. Benchtop scale experimental setups, such as in \cite{Caravaca:2016fjg,Caravaca:2016ryf,Gruszko:2018gzr,Guo:2018kcp,Li:2015phc}, use the timing and directionality to identify the Cherenkov light.

\begin{figure}[b!]
\centering 
\includegraphics[width=0.45\textwidth]{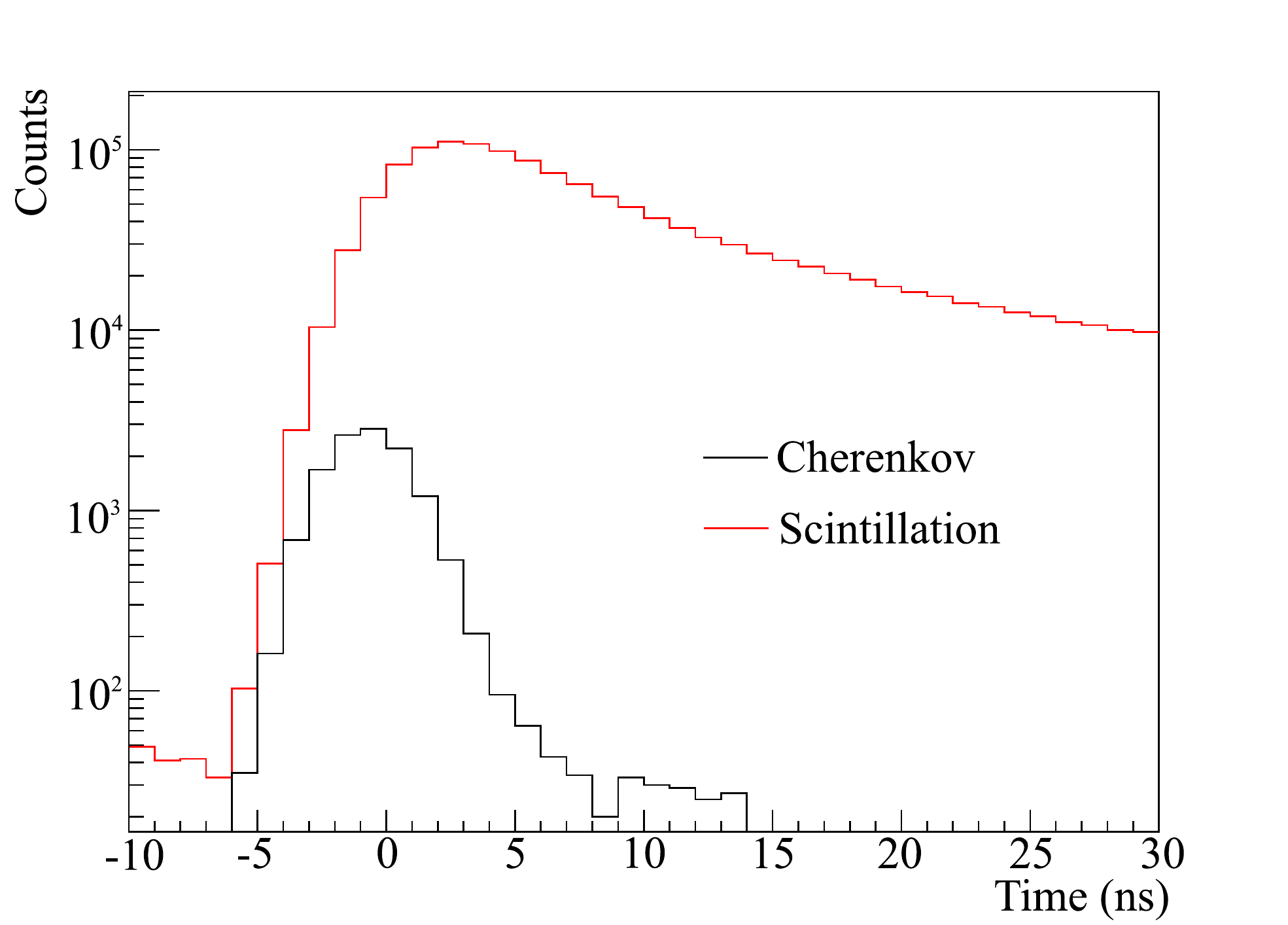}
\caption{The time profile of the detected light for simulated 2.5 MeV electrons at the center of a SNO+-like detector, consisting of about 9000 PMTs with transit time spreads of 1.4 ns, a 6 meter radius acrylic vessel, and about 50\% coverage. The Cherenkov light arrives promptly, but is difficult to identify due to the intrinsic resolution of the photodetectors and high light yield of the scintillator.}
\label{fig:time-residuals}
\end{figure}

Separating the two components in current generation large-scale scintillation-based neutrino detectors is nevertheless very difficult. The transit time spread (TTS) of PMTs is generally around 1.5~ns or larger, making it difficult to resolve the early Cherenkov light. An illustration of the typical timing spectra of the detected light for a SNO+-like detector using liquid scintillator is shown in Figure \ref{fig:time-residuals}, generated using the \texttt{RAT-PAC} software \cite{ratpac}. Even with using both the timing and spatial distributions of the hits, no current generation large-scale scintillator detector has been able to demonstrate the detection of Cherenkov light.

The challenge in discriminating Cherenkov and scintillation light by photon wavelength in large-scale detectors is doing so while maintaining the high detected light yield needed for a low-energy physics program or precision reconstruction and particle ID.  Using water-based liquid scintillator~\cite{YEH201151}, for example, increases the ratio of Cherenkov to scintillation light by reducing the total scintillation light. Using a scintillator like linear alkyl benzene (LAB) with only a small amount of fluor can also be done to slow down the scintillation time profile~\cite{JinpingNeutrinoExperimentgroup:2016nol} and then timing can be used to identify Cherenkov light; however, this again comes with a consequent reduction in scintillation light yield. Adopting a simple filtering scheme, or using sets of photon sensors of different wavelength sensitivities~\cite{Aberle:2013jba}, also reduces total light yield because the detection area taken up by filtered photon sensors can only be used for one photon wavelength band. What is needed is a way to {\it sort} photons by wavelength, directing different wavelength bands toward relevant photon sensors, and doing this in a way that loses as little timing or position information as possible.

In earlier work~\cite{Kaptanoglu:2018sus} we showed that sorting by wavelength can be done using dichroic reflectors, and that broadband (falling as $1/\lambda^2$) Cherenkov light can be distinguished from narrow-band scintillation light, in LAB scintillator doped with 2,5-Diphenyloxazole (PPO). To turn this approach into something that could be used in a large-scale detector, we have configured the dichroic filters into a Winston-style light concentrator, the ``dichroicon.'' As is well known, Winston cones provide optimal light collection for non-imaging detectors \cite{Winston:71}, and have been used in other large-scale neutrino detectors~\cite{Boger:1999bb,Alimonti:2008gc}. An additional advantage to a large-scale detector of using long-wavelength photons to identify Cherenkov light is that the long-wavelength photons travel faster and are scattered and absorbed far less than short-wavelength photons \cite{Wurm:2010ad}, thus preserving more of the directional information of the Cherenkov light. 

\begin{figure}[b!]
\centering 
\includegraphics[width=0.45\textwidth]{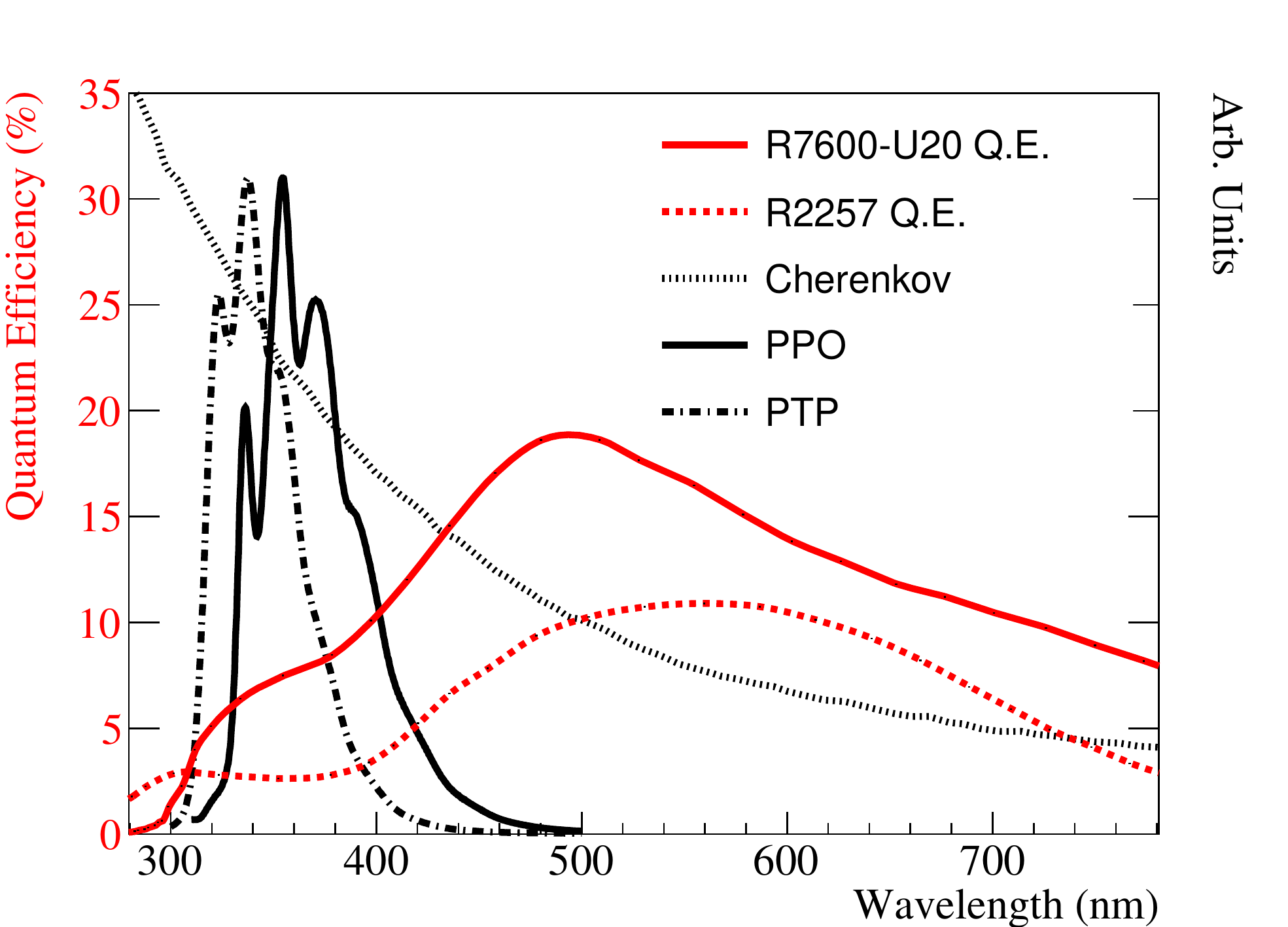}
\caption{The quantum efficiency of the PMTs used in the various measurements compared to the Cherenkov emission spectrum and the emission spectra of the fluors PPO and PTP. The three emission spectra are arbitrarily scaled and show shape only. The quantum efficiency curves for the R7600-U20, R2257, and R1408 are taken from \cite{r7600u20,r2257,Biller:1999ik} respectively. The PPO and PTP emission spectra were taken from the PhotochemCAD database \cite{photochemCAD}.}
\label{fig:wavelengths}
\end{figure}

In this context, this paper refers to photons with wavelengths between around 450 to 900~nm as `long-wavelength' and photons between 350 to 450~nm as `short-wavelength'. The shortest wavelength photons between 300 to 350~nm are absorbed by scintillator and re-emitted at longer wavelengths. The emission spectra of common fluors such as PPO, as shown in Figure \ref{fig:wavelengths}, peak around 360 to 380~nm and tail off by 450~nm, leaving primarily Cherenkov light emission above this wavelength.  Red-sensitive PMTs can be used to detect this long-wavelength light and have quantum efficiencies that can extend to around 800 to 900~nm, also shown in Figure \ref{fig:wavelengths}. 

\begin{figure}[b!]
\centering 
\includegraphics[trim=0.2cm 0.2cm 0.2cm 0.2cm, clip=true, width=0.45\textwidth]{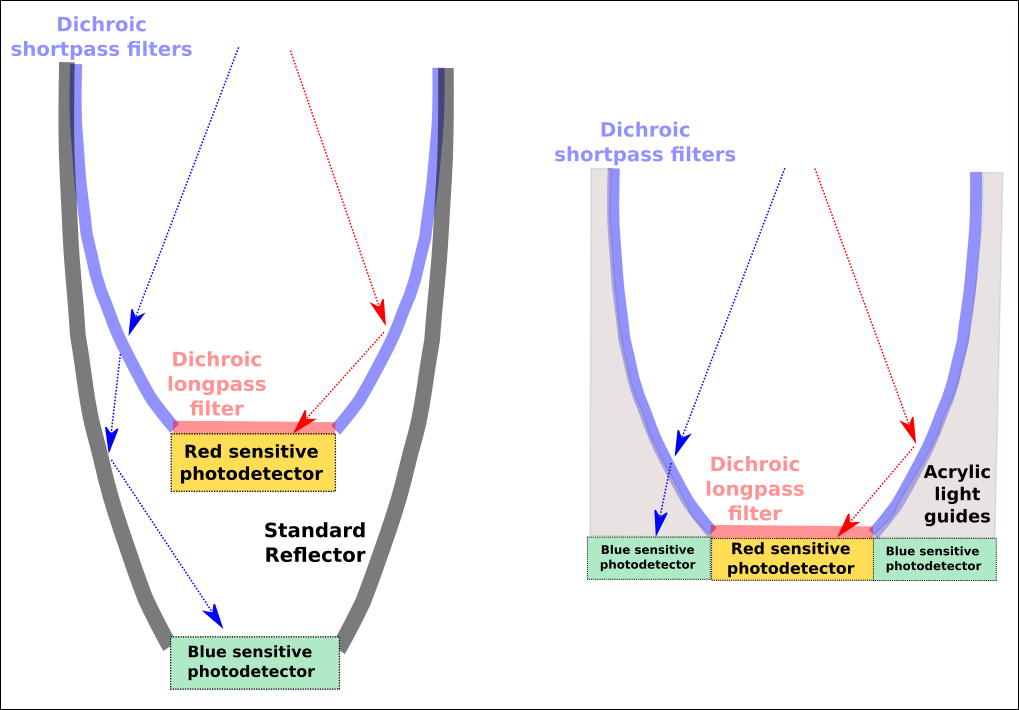}
\caption{Simple schematics for two potential options for a system to detect light sorted by the dichroicon. On the left shows an option where a parabolic reflector is built around the dichroicon to detect the short-wavelength light. On the right an option using acrylic light guides to direct the short-wavelength light to one or more photodetectors. In this case, a pixelated light detector such as a large area picosecond photodetector (LAPPD) might be an ideal sensor for the dichroicon. In both designs the long-wavelength light is detected at the aperture of the dichroicon. Neither of these full designs are constructed in this paper, but are included to show potential detection schemes for the short- and long- wavelength light. The device built for this paper is shown in Figure \ref{fig:setup-schematic}. The blue and red lines show possible photon tracks for short- and long-wavelength light respectively.}
\label{fig:schematic}
\end{figure}

The dichroicon follows the off-axis parabolic design of an ideal Winston cone but is built as a tiled set of dichroic filters and does not achieve the idealized shape. The filters are used to direct long-wavelength light towards a central red-sensitive PMT, while transmitting the shorter wavelength light through the `barrel' of the Winston cone to secondary photodetectors. This is possible because of the remarkable property of the dichroic reflectors, which reflect one passband of light (below or above a `cut-on' wavelength) while transmitting its complement, with very little absorption. As shown schematically for two possible designs in Figure \ref{fig:schematic}, the `barrel' of the dichroicon is built from shortpass dichroic filters and a longpass dichroic filter is placed at the aperture of the dichroicon. The shortpass filter passes short-wavelength light while reflecting long-wavelength light; the longpass has the complementary response.

The parameter space for optimization of the dichroicon is large, and we have explored only a small fraction of it in the work presented here.  The features of the dichroicon that can be varied include: length and geometric field of view, cut-on wavelength for different dichroic filters (they need not all be the same across the cone), photon sensor type and response for both short- and long-wavelength bands, presence or absence of additional filtering at the aperture, shape and reflectivity of the lightguide, and size and configuration of the photon sensors used to detect both sets of light. In fact, a multi-band dichroicon could be designed, which simply nests various dichroicons within each other, if more than two passbands were needed.  Our design here has been constrained by available PMT sizes and sensitivities, and sizes and shapes of available dichroic filters. For a dichroicon that would be deployed in a real detector, the optimization would depend upon the physics goals, the target material and any added fluors, and the fiducial volume of the detector.  

\begin{figure}[b!]
\centering 
\includegraphics[trim=25cm 10cm 40cm 12cm, clip=true, angle=270, width=0.45\textwidth]{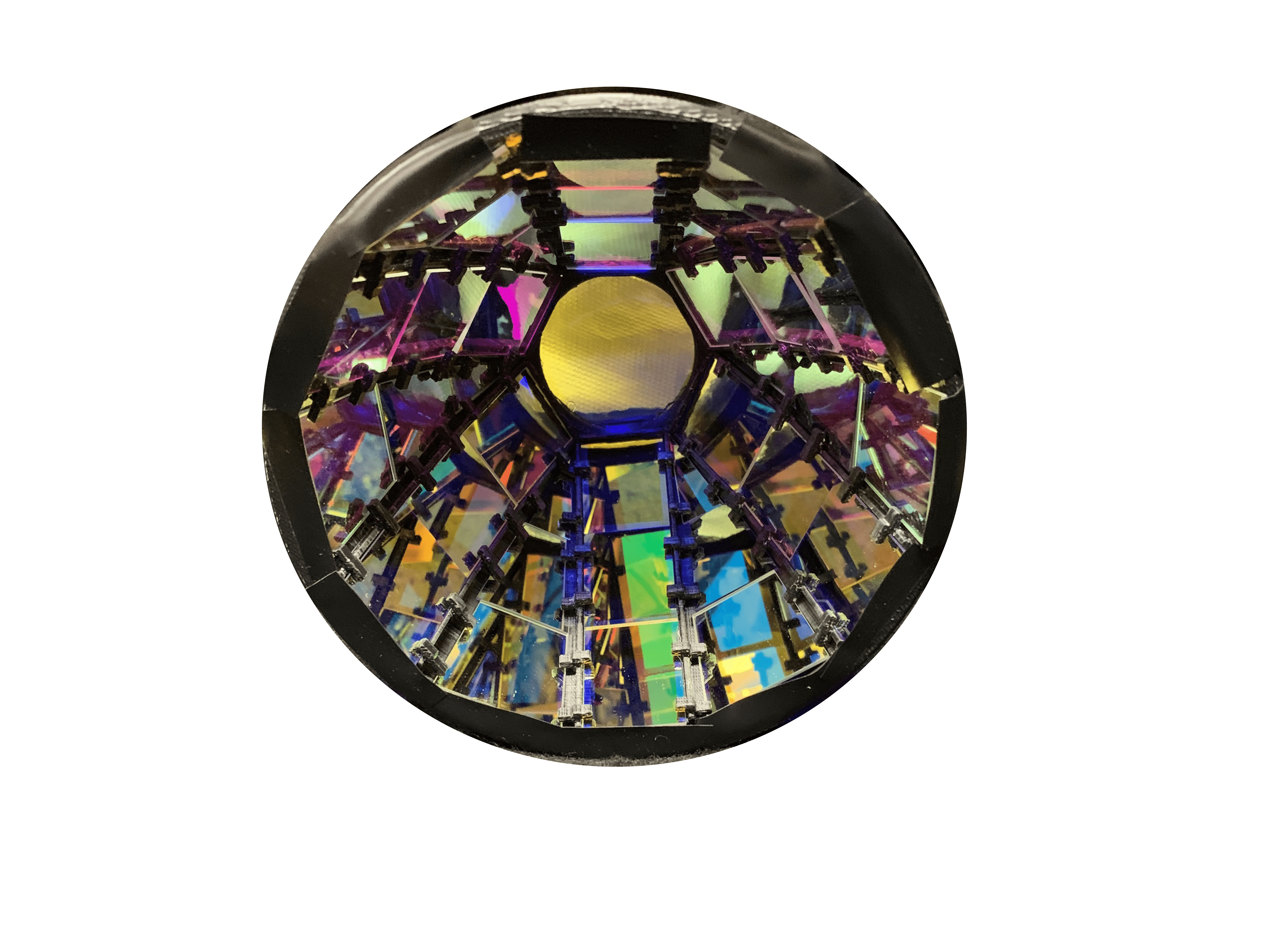}
\caption{A head-on view of the dichroicon. The filters are held in a custom 3D printed plastic holder and can be easily swapped. The shortpass filters tile the barrel of the Winston cone and a central longpass filter is placed at the aperture. A small amount of black electrical tape is used to block a small gap between the filters and the holder at the top of the dichroicon. The outer diameter of the dichroicon is about 150~mm and the inner diameter, where the long-pass filter is located, is about 50~mm.}
\label{fig:dichroicon}
\end{figure}

In our design, a 3D printed structure is tiled with dichroic filters, as shown in Figure \ref{fig:dichroicon}. Around the barrel of the Winston cone are eighteen shortpass filters from Edmund Optics \cite{edmund} and eighteen from Knight Optical \cite{knight}, the latter of which are custom cut to trapezoidal shapes to fill the surface area. A longpass filter from Knight Optical, 50~mm in diameter, sits at the center of the dichroicon. A second dichroicon is also constructed using several different filters. The specific filters used for the two dichroicons are presented in Tables \ref{tab:dichroicon1} and \ref{tab:dichroicon2} respectively. The dichroicon pictured in Figure \ref{fig:dichroicon} and presented in Table \ref{tab:dichroicon1} is used for the majority of the measurements in this paper, and is referred to simply as `the dichroicon'. The measurements with the dichroicon were done with both a Cherenkov source and with a LAB-PPO source. The second dichroicon is used primarily for measurements in LAB with p-Terphenyl (PTP) as the fluor, which has a shorter wavelength emission spectrum than PPO. We refer to that device as dichroicon-2, and the only difference from the dichroicon is the cut-on wavelength of the rectangular shaped short-pass filters used in the barrel of the device is slightly reduced and the cut-on wavelength of the central longpass filter used at the aperture is slightly reduced. All measurements in this paper are made with the dichroicon (rather than the dichroicon-2) other than those presented in Section \ref{sec:results-labptp}.

\begin{table*}[t!]
    \caption{The details for the filters used for the dichroicon shown in Figure \ref{fig:dichroicon}. The cut-on wavelength is given for an average incidence angle of 45$^{\circ}$.}
    \label{tab:dichroicon1}
    \centering
    \begin{tabular}{ccccccc} 
         \hline\hline\noalign{\smallskip}
          Cut-On (nm) & Pass & Shape & Dimensions (mm) & Quantity & Manufacturer & Online Datasheet \\
         \noalign{\smallskip}\hline\noalign{\smallskip}
         500 & Short &   Rectangular   & 25.2 $\times$ 35.6 & 18 & Edmund Optics & \cite{edmund} \\
         453 & Short &   Trapezoidal   & 35 $\times$ 35 $\times$ 35 $\times$ 25 & 6 & Knight Optical & \cite{knight} \\  
         453 & Short & Trapezoidal & 25 $\times$ 35 $\times$ 35 $\times$ 14 & 6 & Knight Optical & \cite{knight} \\
         453 & Short & Triangular & 14 $\times$ 35 $\times$ 35 & 6 & Knight Optical & \cite{knight} \\ 
         480 & Long & Circular & $\boldsymbol{\oslash}$ 50 & 1 & Knight Optical & \cite{knightlp} \\ 
         \noalign{\smallskip}\hline\hline 
    \end{tabular}
\end{table*}

\begin{table*}[t!]
    \caption{The details for the filters used for the dichroicon-2. The cut-on wavelength is given for an average incidence angle of 45$^{\circ}$.}
    \label{tab:dichroicon2}
    \centering 
    \begin{tabular}{ccccccc}
         \hline\hline\noalign{\smallskip}
         Cut-On (nm) & Pass & Shape & Dimensions (mm) & Quantity & Manufacturer & Online Datasheet \\
         \noalign{\smallskip}\hline\noalign{\smallskip}
         450 & Short & Rectangular & 25.2 $\times$ 35.6 & 18 & Edmund Optics & \cite{edmund450} \\
         453 & Short & Trapezoidal & 35 $\times$ 35 $\times$ 35 $\times$ 25 & 6 & Knight Optical & \cite{knight} \\
         453 & Short & Trapezoidal & 35 $\times$ 35 $\times$ 35 $\times$ 14 & 6 & Knight Optical & \cite{knight} \\
         453 & Short & Triangular & 14 $\times$ 35 $\times$ 35 & 6 & Knight Optical & \cite{knight} \\
         462 & Long & Circular & $\boldsymbol{\oslash}$ 50 & 1 & Knight Optical & \cite{knightlp480} \\ 
         \noalign{\smallskip}\hline\hline 
    \end{tabular}
\end{table*}

The cut-on wavelength specified in Tables \ref{tab:dichroicon1} and \ref{tab:dichroicon2} refers to wavelength corresponding to the 50\% transmission crossing, as the filter either goes from transmitting to reflecting or vice versa. The choice for the cut-on, manufacturer, size, and shape of the filters was motivated primarily by availability, cost, and design limitations. The choice to use cut-on wavelengths between 450 to 500~nm for the dichroicon was motivated largely by the result in \cite{Kaptanoglu:2018sus}, where measurements made with LAB+PPO and a single 500~nm dichroic filter showed excellent performance. Shorter pass filters were used for the dichroicon-2, the reasons for which are detailed in Section \ref{sec:results-labptp}.

This paper focuses on detecting the light at both the aperture and barrel of the Winston cone using PMTs. There are, however, many possible options for photon sensing. One interesting option is LAPPDs, which provide excellent time resolution of around 50 ps \cite{Wetstein:2012qxa} over their active areas, which consist of 350 square centimeter pixels. By coupling an LAPPD to a dichroicon one could use the central pixels to detect the long-wavelength light and the outer pixels for the short-wavelength light. Also possible instead of PMTs is an array of silicon photomultipliers (SiPMs), similar to the device described in \cite{LI2019162334}, to perform pixelated detection of the light sorted by the dichroicon.

The method for detecting the short-wavelength light is complicated by the fact that the entire barrel of the Winston should be instrumented to provide maximal collection efficiency. Placing PMTs directly behind the barrel of the cone would be an expensive and inefficient way to collect the light, although it would preserve the timing of the photons. There are many ideas for better ways to collect these photons -- one could use acrylic light guides to direct the light back towards one or several blue sensitive PMTs. Another idea is to use a second parabolic reflector, built from reflecting material instead of filters, that wraps around the dichroicon, more efficiently directing the short-wavelength light toward the photon sensor. A simpler option consists of a cylinder that is lined with reflective material or paint, which reflects the short-wavelength photons back to a single PMT. This final option was chosen for our experimental setup, as will be discussed in Section \ref{sec:experimentalsetup}.

All of these options will slightly degrade the timing of the short-wavelength photons by reflecting the photons one or more times before they are detected. Particularly for scintillation detectors, where the timing of the photons is already spread out by the intrinsic emission spectrum of the scintillator, this will be a small effect, and it is part of the reason for choosing to tile the barrel of the dichroicon with shortpass filters, rather than building the complementary design. 

Although we have chosen to include a longpass dichroic filter at the dichroicon aperture, its presence ultimately depends on the physics goals of an experiment and the configuration of the photon sensors.  The advantage of a longpass dichroic filter at the aperture rather than a simple longpass absorbing filter, is that short-wavelength light that hits the aperture is reflected rather than absorbed, and thus total short-wavelength light yield is not affected. The majority of this light, in our configuration, would have to be detected by another device---for example, by PMTs on the other side of a large detector. A different optimization of the dichroicon, however, might lead to a different choice for the aperture filter.

Nevertheless, even with a longpass dichroic filter at the aperture of the dichroicon, we find that some short-wavelength light does leak through. For Cherenkov/scintillation separation, even this small amount of leakage is noticeable, because there is so much more scintillation than Cherenkov light generated at the source. Thus we have included a longpass absorbing filter {\it behind} the longpass dichroic filter at the dichroicon aperture in some of our measurements, which leads to improved purity of Cherenkov photons detected there. While this means that some of the short-wavelength light is lost from the system, it is a tiny amount, as it is only the small number that leak through the longpass dichroic filter that are absorbed. A photon sensor with better timing at the aperture would likely make the longpass absorbing filter unnecessary for Cherenkov/scintillation separation, as the small amount of scintillation light that leaks through could be distinguished by timing.

Our primary goal in the measurements of this dichroicon is to demonstrate the ability to sort photons for both a Cherenkov and scintillation source. Our measurements with a Cherenkov source are designed as a demonstration that the dichroicon works as intended, in an easy to study system, with application for large-scale neutrino detectors such as Hyper-Kamiokande. Our measurements with scintillation sources will further demonstrate the photon sorting technique in addition to providing a way to separate Cherenkov and scintillation light. 

In Section \ref{sec:filter-characterization} we discuss the experimental setup and results of our dichroic filter characterization, which provides critical input into our simulation software. In Section \ref{sec:pmtcal} the calibration of the PMTs used in the dichroicon setup is discussed, which is necessary for a quantitative understanding of the dichroicon results. In Section \ref{sec:dichroicon_measurements} the benchtop setup, data analysis, and results for the dichroicon measurements are presented. These results include measurements using a Cherenkov source and two different scintillation sources, a variety of different dichroic filters and absorbing longpass filters, and two different red-sensitive PMTs at the aperture of the dichroicon. 

Finally, it should be noted that dichroic filters have appeared recently in several other potential photon detection devices for large-scale neutrino detectors. These filters are starting to be studied in more detail for use as a photon-trap device for Hyper-Kamiokande \cite{Rott:2017lip} and for the ARAPUCA and X-ARAPUCA light trap designs for ProtoDUNE and DUNE \cite{Machado:2016jqe}.

\section{Dichroic Filter Characterization}\label{sec:filter-characterization}

Manufacturers typically provide data for one or two incidence angles for dichroic filters, scanned over wavelength \cite{edmund,knight}; however, dichroic behavior as a function of incidence angle, which is needed for simulation studies and optimization, is not provided.  We have therefore characterized them on the benchtop in two ways: first using LED sources and well-understood PMTs and second using LED sources and a spectrometer. The measurement using PMTs provides both transmissivity and reflectivity information, but the data is averaged over the spectrum of the LED. The spectrometer data provides only transmissivity, but the detailed wavelength dependence of the filter can be extracted. We have also measured transmissivity of the filter in water.

\subsection{Experimental Setup}\label{sec:filtercharacterization-setup}

\begin{figure}[b!]
    \centering
    \includegraphics[trim=0.2cm 0.2cm 0.2cm 0.2cm, clip=true, width=0.45\textwidth]{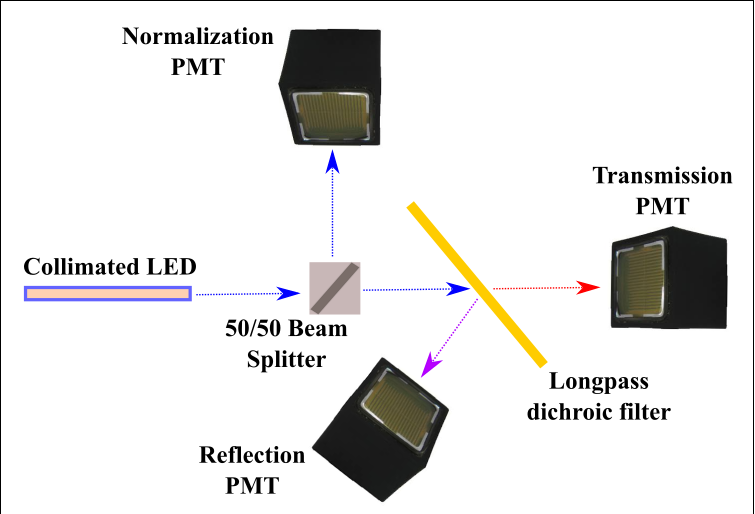}
    \caption{The dichroic filter characterization setup. The normalization, transmission, and reflection PMTs are R7600-U200 PMTs. The dichroic filter is held on a rotating stage.}
    \label{fig:filter-characterization-setup}
\end{figure}

A schematic of the dichroic filter characterization experimental setup using PMTs is shown in Figure \ref{fig:filter-characterization-setup}. In this setup, a collimated LED is directed toward a 50/50 beamsplitter \cite{beamsplitter}. One output of that beamsplitter goes toward an R7600-U200 PMT, referred to as the normalization PMT, which provides a measure of the LED intensity, which changes slightly across datasets. The other output goes to a dichroic filter, which both transmits and reflects the incoming photons. The transmitted and reflected light is detected by R7600-U200 PMTs, referred to as the transmission and reflection PMTs respectively. The three PMTs are operated at -800~V.

The dichroic filter is located on a rotating stage and the angle can be chosen in 1$^{\circ}$ increments. We measured the response at incidence angles from $\sim$ 0 - 60$^{\circ}$, where 0$^{\circ}$ indicates light impinging normal to the surface. Given the geometry of the setup, most significantly the opening angle of the collimated LED beam, very large incidence angles $>$ 60$^{\circ}$ were not possible to measure. Additionally, measurements of reflection were not made at incident angles less than 15$^{\circ}$ because of shadowing by parts of the setup of the reflected light.

LEDs at wavelength of 385, 405, 450, 505, 555, 590, and 630~nm from Thorlabs were used to probe the filter response across wavelength. The data sheets are available online \cite{thorlabs}. The spectral FWHM of the LEDs range from 12 to 30~nm and no filters were used to narrow the wavelength range of the beam.

The LEDs are pulsed with 40~ns wide 3~V square pulses at 1 kHz. This output is split, one side is used to trigger the oscilloscope acquisition and the other goes to the LED. At these settings the LED output provides a relatively high intensity source, resulting in the collection of around 100 photoelectrons (PEs) per triggered event at the normalization PMT. In general, due to the nature of the dichroic filter, either the reflection or transmission PMT views a similar number of PE, while the other detects very few photons over the entire dataset. 

\subsection{DAQ and Data Analysis}\label{sec:data-analysis-1}

The data acquisition (DAQ) system is a Lecroy WaveRunner 606Zi 600 MHz oscilloscope which digitizes the analog signals from the PMTs. The data is sampled every 100~ps in 200~ns long waveforms. The oscilloscope has an 8-bit ADC with a variable dynamic range, which allows for roughly 100 $\mu$V resolution. The \texttt{LeCrunch} software \cite{latorre} is used to read out the data, formatted in custom \texttt{hdf5} files, over ethernet connection.

\texttt{C++}-based analysis code runs over the \texttt{hdf5} files to calculate the amount of light collected by the normalization, transmission, and reflection PMTs. Each PMT signal is integrated to produce a charge. The gain of the PMTs is set such that if a single photon is detected the peak of the charge distribution sits around 1 pC. For each triggered event, the charge is converted to number of photons detected, which is summed over the total data set.

\subsection{Results}\label{sec:filtercharacterization-results}

For each LED, two calibration datasets are taken with no dichroic filter -- one with the LED directed at the transmission PMT and the other with it directed toward the reflection PMT (still including the beamsplitter and normalization PMT). These datasets are used to measure both PMT responses under the condition where no dichroic filter is blocking the LED output and are used to normalize to an expected intensity for 100\% transmission or reflection.

The calculated transmission $T$ through the dichroic filter is given below in Equation \ref{eq:transmission}: 
\begin{equation}\label{eq:transmission}
T = \frac{T_{F}}{T_{NF}} \times \frac{N_{NF}}{N_{F}}
\end{equation}
The first term, $T_{F}$, is the total amount of light detected at the transmission PMT and is divided by the total amount of light detected by the transmission PMT when no dichroic filter was present, $T_{NF}$. This gives the fractional transmittance of the filter, under the assumption that the intensity of the LED did not change. To correct for realistic variations in the LED intensity, we multiply by a second term, which is the normalization PMT measurement of the LED intensity during the data taking with no filter, $N_{NF}$, divided by the normalization PMT measurement during data taking with the filter, $N_{F}$. This second term provides us with the relative change in intensity of the LED between the two measurements. The same equation is used to calculate the reflectance, $R$, where the data for the reflection PMT is used instead: 
\begin{equation}\label{eq:reflection}
R = \frac{R_{F}}{R_{NF}} \times \frac{N_{NF}}{N_{F}}.
\end{equation}

\begin{figure}[t!]
    \centering
    \includegraphics[width=0.45\textwidth]{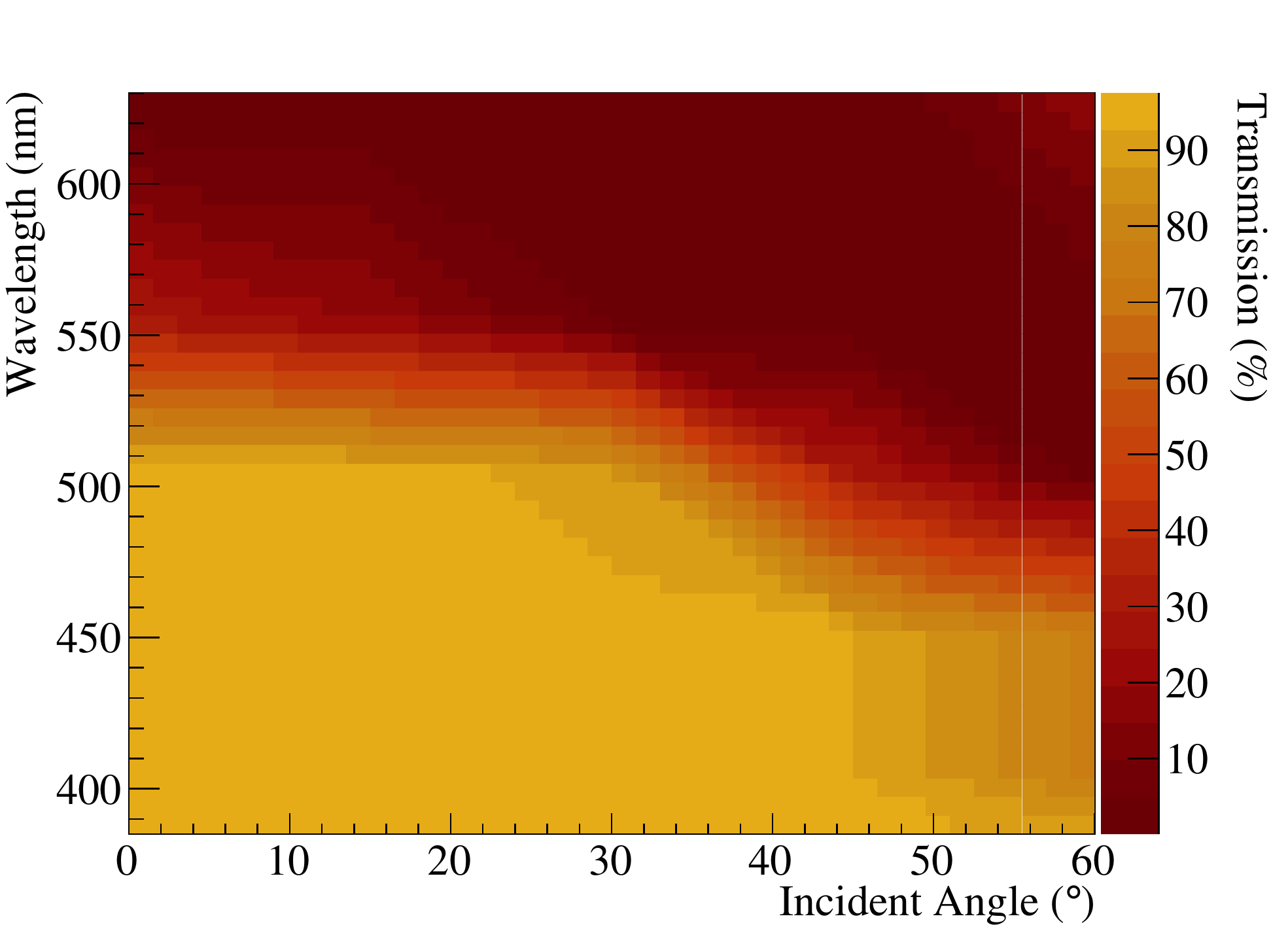}
    \caption{The transmission data for the 500~nm shortpass filter for incidence angles between 0 to 60$^\circ$. The percent transmission was calculated using Equation \ref{eq:transmission}.}
    \label{fig:shortpass-transmission}
\end{figure}

\begin{figure}[b!]
    \centering
    \includegraphics[width=0.45\textwidth]{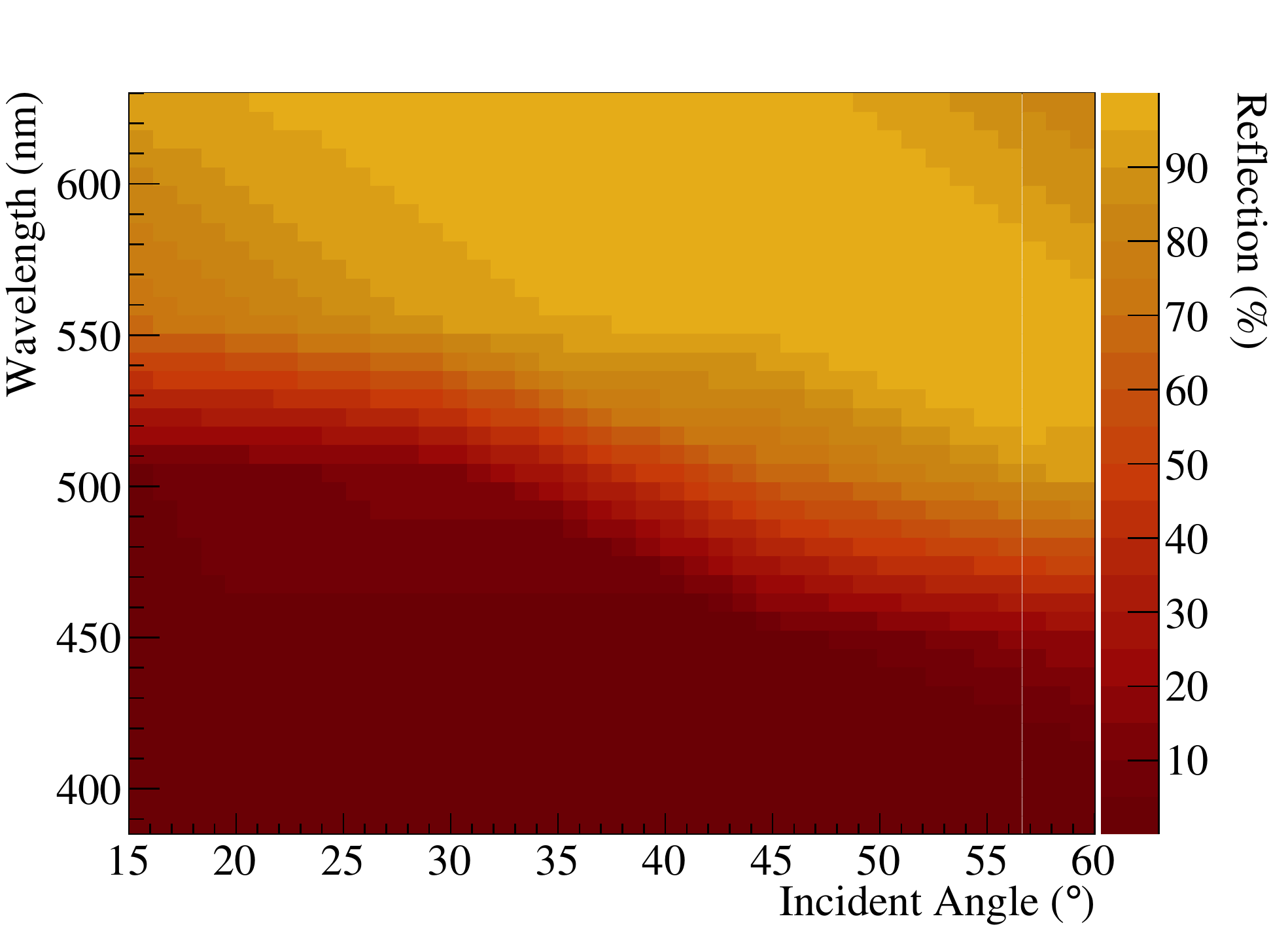}
    \caption{The reflection data for the 500~nm shortpass filter for incidence angles between 15 to 60$^\circ$. Note that the reflected data only extends to incident angles of 15$^{^\circ}$ due to shadowing effects in the setup. The percent reflection was calculated using Equation \ref{eq:reflection}.}
    \label{fig:shortpass-reflection}
\end{figure}

The results for the transmissivity and reflectivity of the 500~nm short-pass filters used in the barrel of the dichroicon are shown in Figure \ref{fig:shortpass-transmission}. and \ref{fig:shortpass-reflection}.

\subsection{Spectrometer Results}
\label{sec:spectrometer-results}

The experimental setup to measure the transmission as a function of wavelength and incidence angle with an Ocean Optics USB-UV-VIS Spectrometer is a simplified version of the setup shown in Figure \ref{fig:filter-characterization-setup}. The PMTs and beam-splitter are removed and the transmitted light is detected with the spectrometer. While this simple setup does not provide reflectivity values, the transmitted data captured is far more detailed in terms of the behavior as a function of wavelength. To span relevant wavelengths between 350 to 750~nm, three different light sources are used: a 365~nm LED, a 405~nm LED, and a white LED which spans 420 to 750~nm. 

\begin{figure}[b!]
    \centering
    \includegraphics[width=0.45\textwidth]{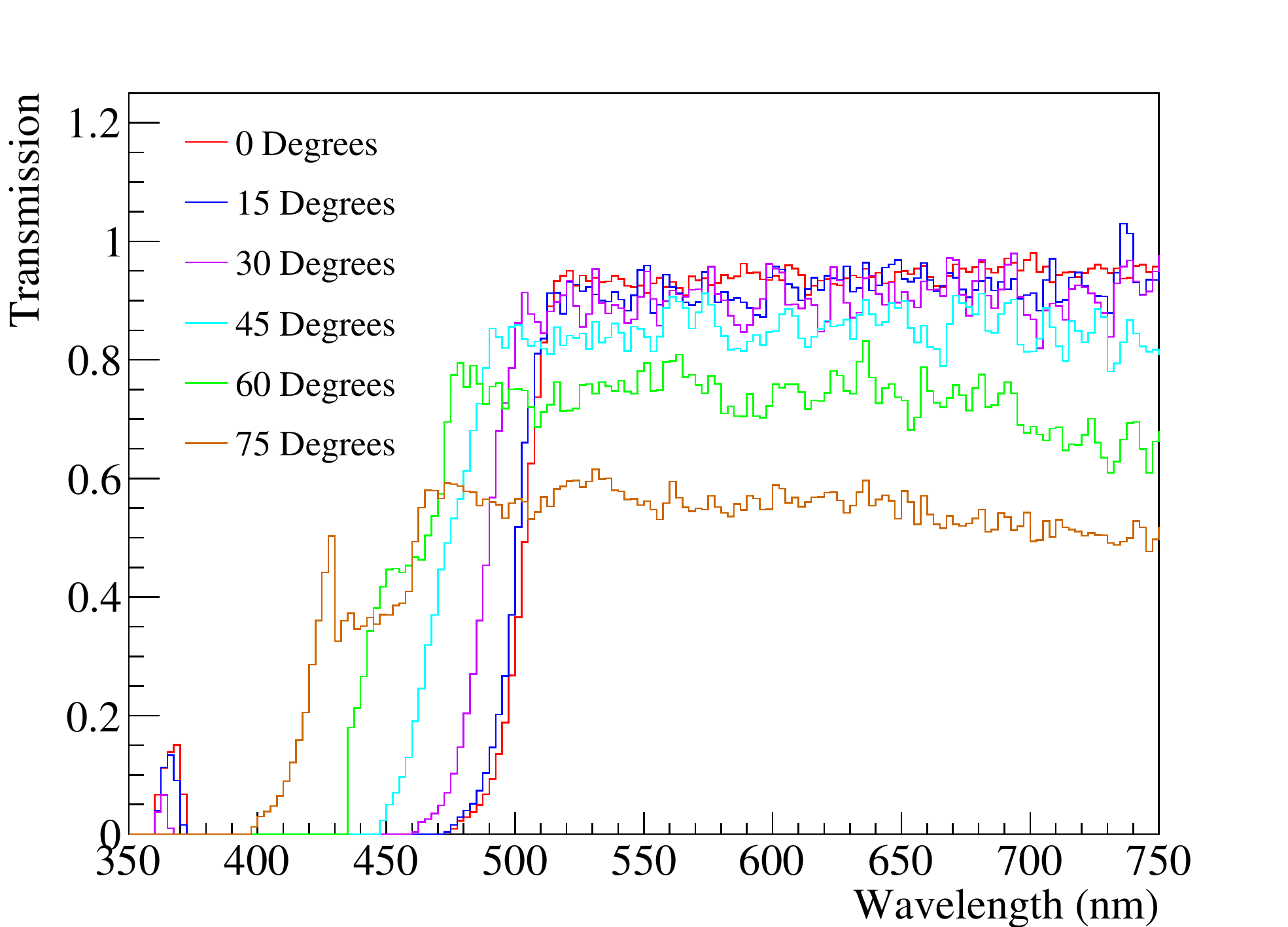}
    \caption{The transmission through the 480~nm longpass filter, used at the aperture of the dichroicon, as a function of wavelength, for multiple incidence angles.}
    \label{fig:transmission-lp-spectrometer}
\end{figure}

Data is taken with no filter to understand the spectrum and intensity of each of the LEDs. The dichroic filter is added between the collimated LED and the spectrometer at varying incidence angles. This data is normalized to the no filter data to calculate the absolute transmission. The resulting transmission is shown in Figure \ref{fig:transmission-lp-spectrometer} for the 480~nm longpass filter that is used at the aperture of the dichroicon.

This data is consistent with the PMT data that was taken for this filter. Perhaps most interestingly, the small amount of short-wavelength leakage through this filter at high incidence angles becomes clear and will be important in understanding the results for the dichroicon. This measurement is performed for all of the different types of filters deployed in the dichroicon, specified in Table \ref{tab:dichroicon1}.

\begin{figure}[t!]
    \centering
    \includegraphics[width=0.45\textwidth]{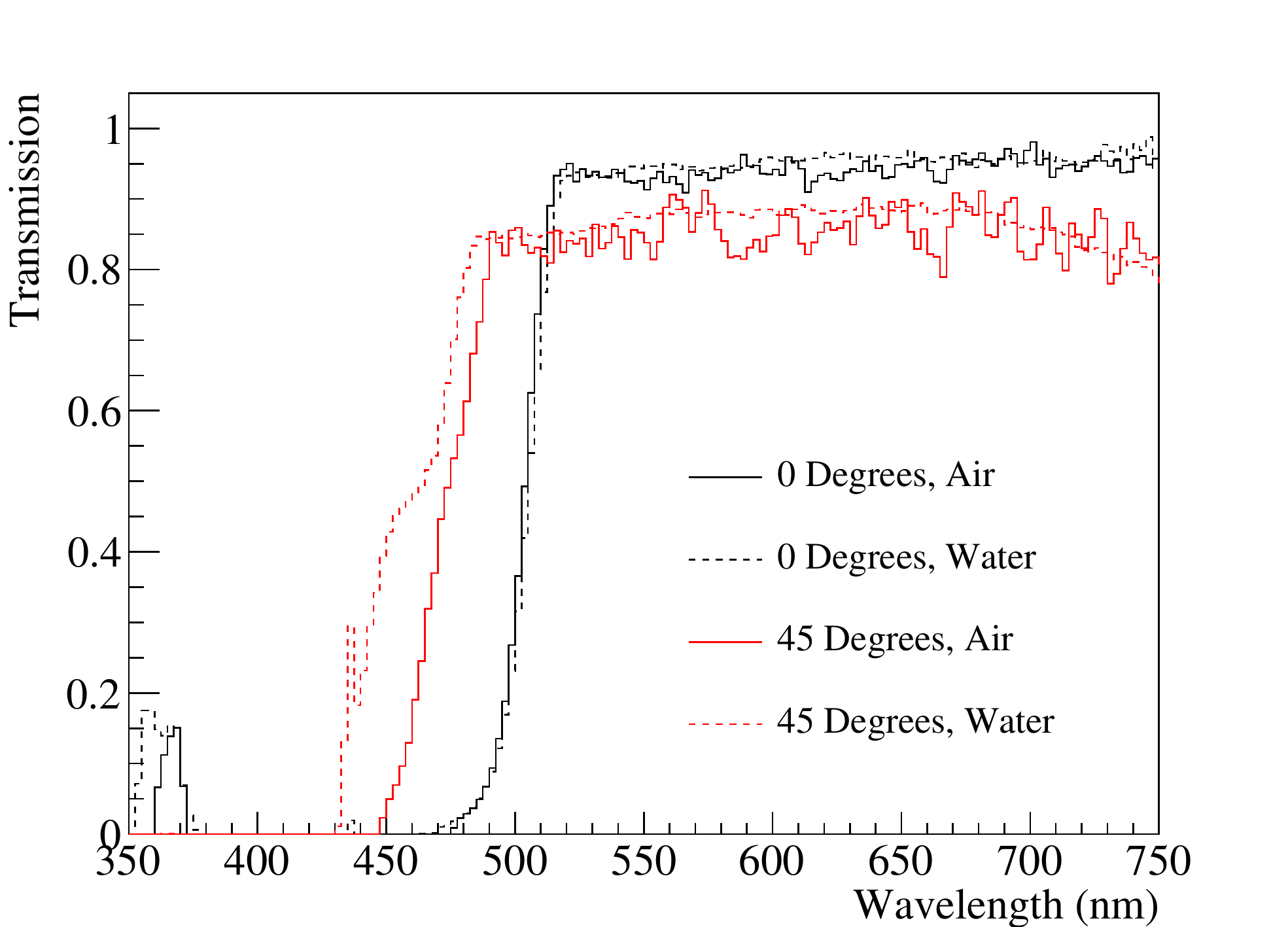}
    \caption{The transmission through the 480~nm longpass filter, used at the aperture of the dichroicon, as a function of wavelength, for 0$^{\circ}$ and 45$^{\circ}$ incidence angles, in air and in water.}
    \label{fig:water-transmission-lp-spectrometer}
\end{figure}

The behavior of the filters will ultimately need to be mapped for the specific fluid that the dichroicon is submerged in. The buffer fluid around the PMTs is commonly water, and measurements with the filter placed in a water bath are made to understand the expected change in performance. The setup and technique is identical to those in air, and the results show that the dichroic filters do perform differently in water. Figure \ref{fig:water-transmission-lp-spectrometer} shows the results for the 480~nm longpass filter at two incidence angles compared to the data taken in air. As expected, there is a small change to the behavior of the filter, particularly where it transitions from reflecting to transmitting at larger incidence angles. We would anticipate larger shifts were the dichroicon submersed in scintillator or oil.

The results described in this section are used as input into the simulation software used to model the dichroicon, described in Section \ref{sec:simulations}. More detailed studies of these dichroic filters, with data taken at more incidence angles will improve our model in the future.

\section{PMT Calibration}\label{sec:pmtcal}

In the dichroicon measurements presented in Section \ref{sec:dichroicon_measurements}, light is detected using three different Hamamatsu PMTs: an R2257, an R7600-U20, and an R1408.  To compare various dichroicon measurements, we need the relative detection efficiencies of the PMTs The total detection efficiency is given by:
\begin{equation}\label{eq:detection-eff}
\rm QE \times \rm CE \times \rm F
\end{equation}
where QE is the quantum efficiency of the PMT, CE is the collection efficiency of the PMT, and F is the front-end efficiency of the PMT. The QE measures the likelihood a photoelectron is created given an incident photon, and depends on the wavelength of the photon. The CE is the probability that a created photoelectron is guided to the dynode stack and multiplied, creating a signal at the anode. The F measures the efficiency for detecting the signal, given that the PMTs pulses might be lost in the noise or fall below the analysis threshold. In principle an additional efficiency factor is needed for absorption of the PMT glass and photocathode, however this factor should be similar between the three PMTs and thus not impact the relative detection efficiencies.

The QE curves are provided by Hamamatsu for the R2257 and R7600 PMTs \cite{r2257, r7600u20}, and the QE of the R1408 has been measured for the SNO collaboration \cite{Biller:1999ik} but the factor CE $\times$ F must be measured for our setup. For this we use an experimental setup very similar to the one described in Section \ref{sec:filtercharacterization-setup}, as well as identical DAQ and analysis software.

The measurement is perform for two LEDs, at peak wavelengths of 505~nm and 590~nm. The LED is collimated and directed toward the 50/50 beamsplitter, one output of which goes towards an R7600-U200 normalization PMT. The other output is directed toward either a 494~nm or 587~nm bandpass filter, which narrows the wavelength spectrum of the LED so that it only spans a small portion of the QE curve. The output of the bandpass filter is detected by one of the three measurement PMTs. The ratio of the number of photoelectrons detected by the measurement PMT and the normalization PMT is calculated and compared between the three PMTs. The difference in this ratio between the PMTs measures the difference in detection efficiencies. The known QE for each PMT at 494~nm and 587~nm can be factored out, giving a measure of the relative $\rm CE \times \rm F$ factors, referred to as R$_{\rm CEF}$. The values of R$_{\rm CEF}$ calculated relative to the R2257 PMT (the least efficient) at both wavelengths are shown in Table \ref{tab:pmtcal}. Note the agreement between the $R_{\rm CEF}$ values at both wavelengths, as expected based on the fact that neither the collection or front-end efficiency should be wavelength dependant.

\begin{table}[t!]
    \caption{The relative efficiencies of the R7600-U20 and R1408 PMTs. The R$_{\bf CEF}$ measures the efficiency ratio between the R2257 PMT and the other two PMTs, after factoring out the expected difference in the quantum efficiencies. This effectively provides the value for the collection and front-end efficiencies of the other two PMTs. This factor is used when comparing expected numbers of detected photons in the dichroicon measurements, which are performed relative between the PMTs.}
    \label{tab:pmtcal}
    \centering
    \begin{tabular}{c|c|c}
        \hline\hline\noalign{\smallskip}
         PMT  & R$_{\bf CEF}$ at 494~nm &  R$_{\bf CEF}$ at 587~nm  \\
         \noalign{\smallskip}\hline\noalign{\smallskip}
         R7600-U20 & 1.81 $\pm$ 0.04 & 1.86 $\pm$ 0.04 \\ 
         R1408 & 2.02 $\pm$ 0.04 & 1.97 $\pm$ 0.04 \\
        \noalign{\smallskip}\hline\hline
    \end{tabular}
\end{table}

\section{Dichroicon Measurements}\label{sec:dichroicon_measurements}

\subsection{Experimental Setup}\label{sec:experimentalsetup}

\begin{figure}[t!]
\centering 
\includegraphics[angle=270, trim=2cm 6cm 32cm 6cm, clip=true,width=0.22\textwidth]{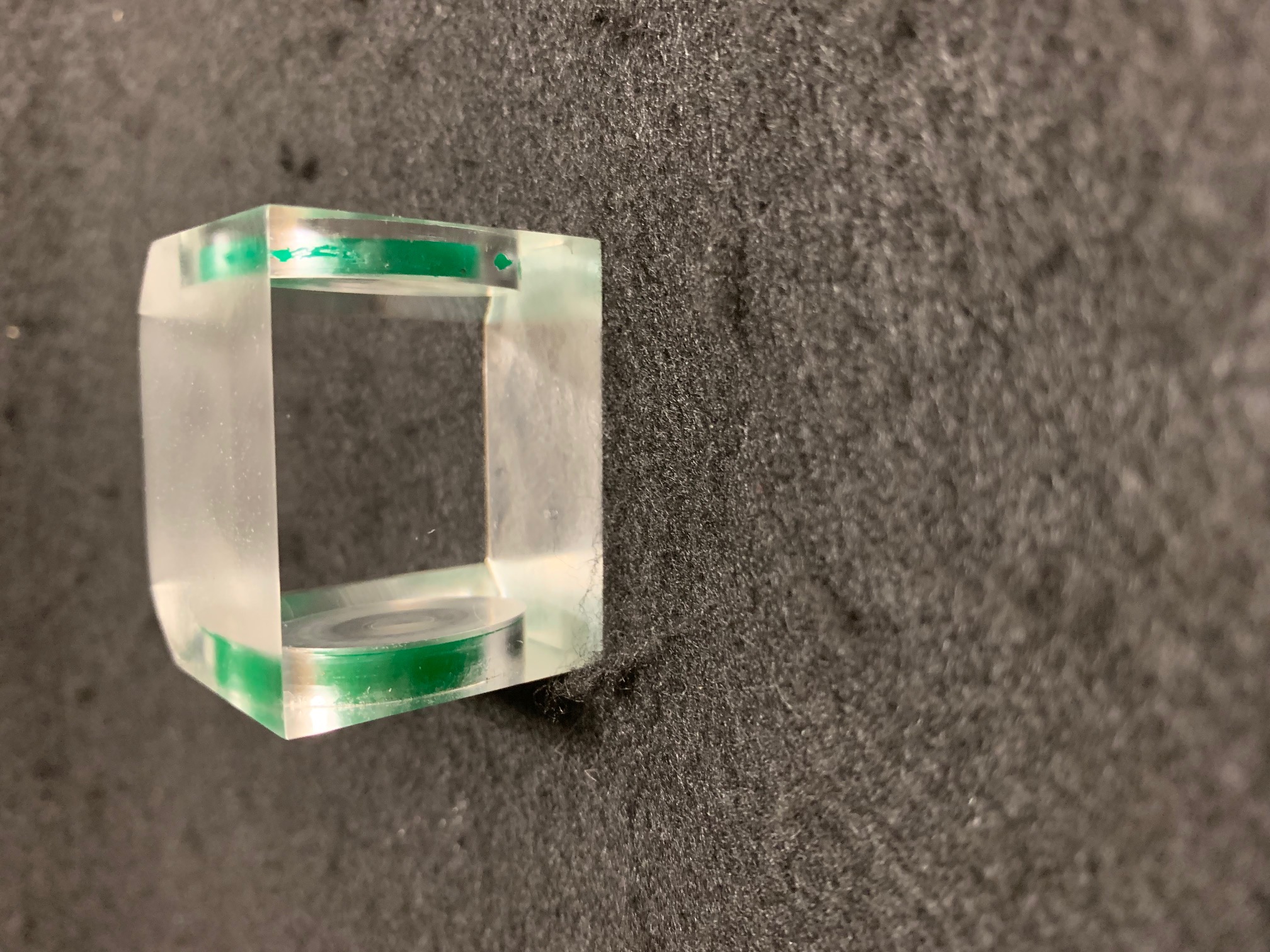}
\includegraphics[angle=270, trim=2cm 6cm 32cm 6cm, clip=true,width=0.22\textwidth]{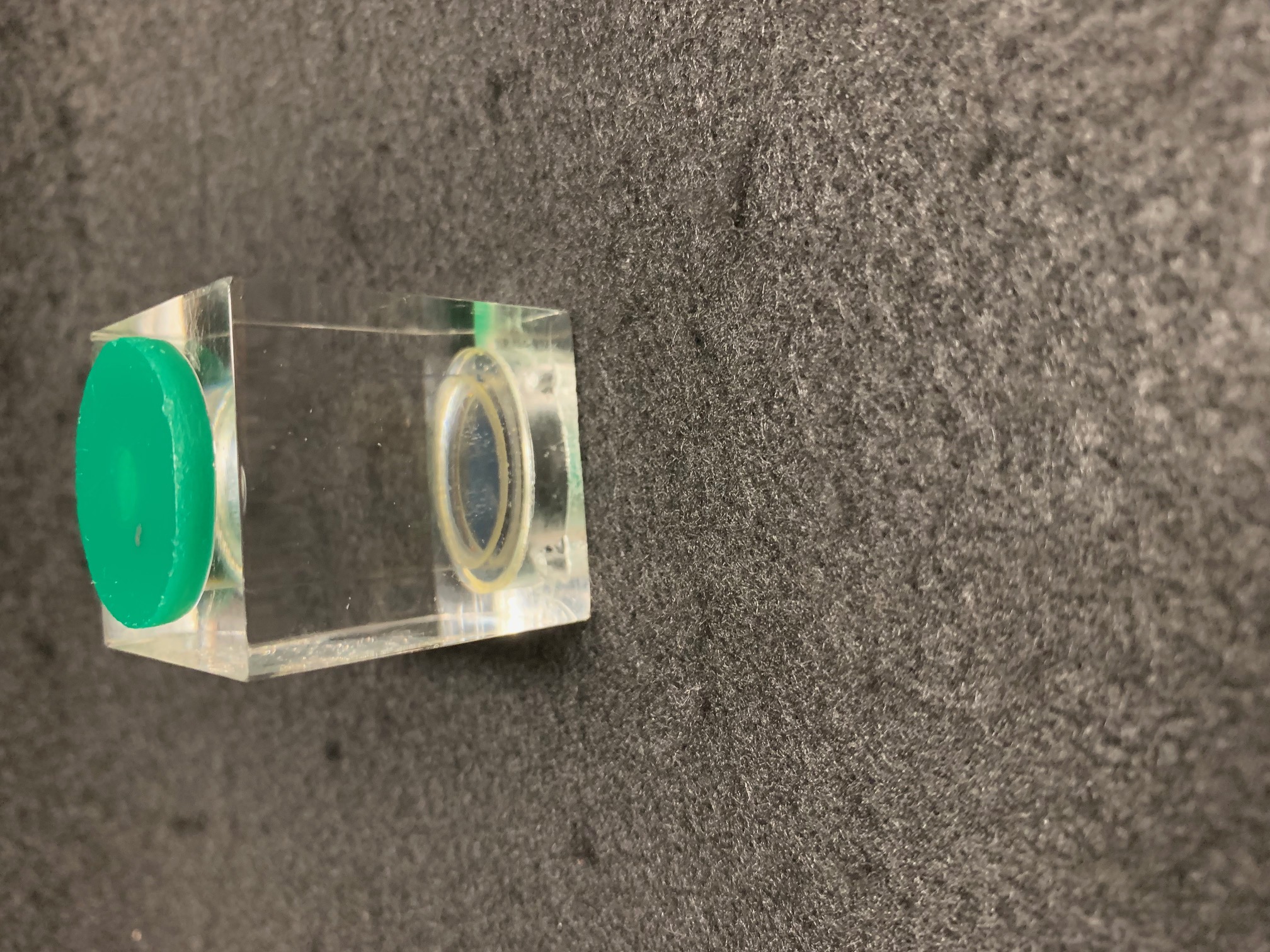}
\caption{The acrylic Cherenkov source (left) and the scintillation source (right). The Cherenkov source is an UVT acrylic block with two $^{90}$Sr sources embedded on its sides. The scintillation source is hollowed UVT acrylic, filled with scintillator, and a $^{90}$Sr source deployed above the scintillator volume. In this picture the scintillation source is filled with LAB+PPO.}
\label{fig:sources}
\end{figure}

The dichroicon experimental setup consists of either a pure source of Cherenkov light or a scintillation source. The Cherenkov source consists of an UV-transparent (UVT) acrylic block embedded with two $^{90}$Sr sources on two sides, shown on the left in Figure \ref{fig:sources}. The scintillation source is a $^{90}$Sr source deployed above a hollowed-out UVT acrylic block filled with scintillator, shown on the right in Figure \ref{fig:sources}. The blocks are 3 x 3 x 3.5 cm cubes. The scintillation source has a machined out cylindrical volume that is 2~cm in diameter. 

The $^{90}$Sr undergoes a 0.546~MeV $\beta^{-}$ decay to $^{90}$Y with a half life of 29.1 years. The $^{90}$Y undergoes a 2.28~MeV $\beta^{-}$ decay to $^{90}$Zr with a half-life of 64 hours. The $\beta^{-}$ particles create either isotropic scintillation light and/or Cherenkov light in the scintillator or acrylic. At these low energies the $\beta^{-}$ particles often undergo several scatters such that the Cherenkov light can be created traveling in any direction.  Our simulation studies of the scintillation source, using the \texttt{GEANT4} implementation in the \texttt{RAT-PAC} simulation package, have shown us that the amount of Cherenkov light created in the acrylic reaching our dichroicon is very small; nearly all photons come from the target scintillator volume in the hollowed-out space of the acrylic block. 

The scintillator of choice is LAB, which has become a popular liquid scintillator due to its ease of handling, high light yield, and compatibility with acrylic. The fluors used in the various measurements are PPO and PTP, both added at a concentration of 2~g/L to the LAB. The primary fluors drastically increase the light output and shift the emission from the UV into the visible, where the PMTs operate most efficiently. Because of its popularity, the LAB+PPO properties have been characterized on the benchtop \cite{OKeeffe:2011dex,Lombardi:2013nla,OSullivan:2012fwt,MarrodanUndagoitia:2009kq,Li2011}. The properties of the fluor PTP are also well-studied \cite{DEVOL1993354}, but for slightly different applications, such as for the X-ARAPUCA devices \cite{Machado:2018rfb}, and it is not a particularly common fluor to dissolve in LAB. Thus, the majority of the measurements in this paper are presented with LAB+PPO.

A Hamamatsu ultra-bialkali R7600-U200 PMT is optically coupled to the acrylic block using Eljen Technology EJ 550 optical grease \cite{eljin}. The PMT acts as a high efficiency fast trigger and provides the time-zero, and is referred to as the trigger PMT. On the other side of the cube is the dichroicon. The filter at the central aperture of the dichroicon is coupled to a PMT using the EJ 550 optical grease. 

Measurements are made separately with two different Hamamatsu PMTs at the aperture: the R7600-U20 and the R2257, which are referred to as the aperture PMTs. The former is a 1-inch (25.4~mm) square PMTs operated at -900V, while the R2257 is a 2-inch (50.8~mm) diameter cylindrical PMT operated at 1500V. These high voltage values were chosen based on recommendations from the Hamamatsu datasheets, and the PMTs were not operating at the same gain. The QE of these PMTs, as well as the emission spectra of the fluors is shown compared to the Cherenkov spectrum in Figure \ref{fig:wavelengths}.

The R7600-U20 has the advantage of a very high efficiency for long-wavelength light, peaking around 20\% at 500~nm, and still at 10\% by 700~nm. Additionally, this PMT has very fast timing, with a measured TTS of around 350~ps. In comparison, the R2257, while relatively efficient at long wavelengths, only peaks around 10\% efficiency close to 600~nm. Additionally, the 900~ps TTS, while still very good, is not as impressive as the R7600-U20. The photocathode area, however, is about five times larger than the R7600-U20, and its cylindrical shape makes it match very well at the center of our dichroicon design.

\begin{figure}[t!]
    \centering 
    \includegraphics[width=0.45\textwidth]{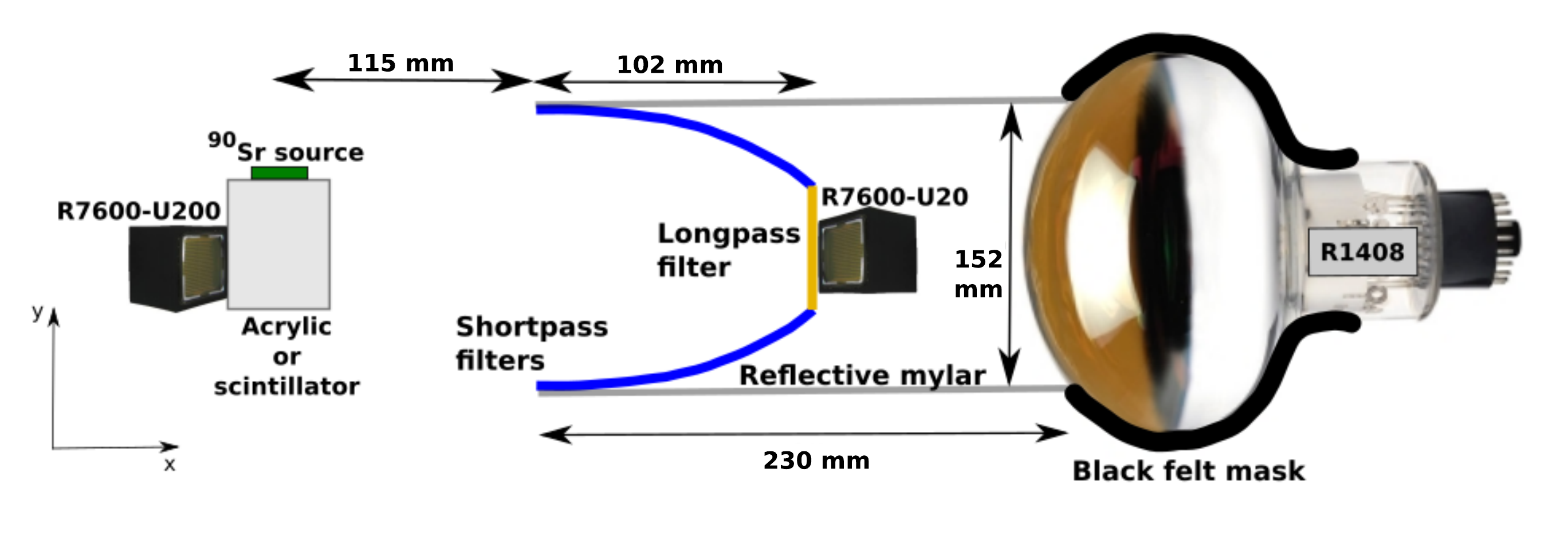}
    \caption{A schematic showing the setup with the dichroicon and reflective cylinder. The R7600-U200 PMT is optically coupled to the acrylic or scintillator source and used as a fast trigger. The long-wavelength light is detected at the aperture of the dichroicon. The short-wavelength light is transmitted through the dichroicon, reflective off of the reflective Mylar lining the cylinder, and detected by an R1408 PMT. The setup with the R2257 aperture PMT is identical to the one shown, except due to the length of the aperture PMT, the reflective cylinder is extended 150~mm. The back of the aperture PMT is covered in reflective foil. The area of the R1408 PMT outside of the reflective cylinder is masked off using felt.}
    \label{fig:setup-schematic}
\end{figure}

\begin{figure}[t!]
    \centering
    \includegraphics[width=0.45\textwidth]{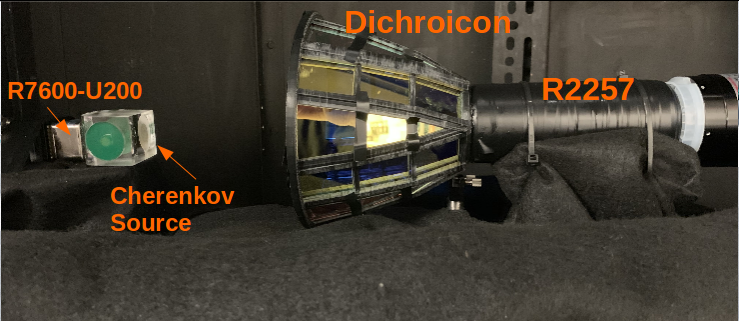}
    \caption{A side view of the dark-box setup with the Cherenkov source. The dichroicon is shown with the R2257 PMT at the aperture. In front of the R2257 PMT is a 480~nm longpass dichroic filter. The barrel of the dichroicon consists of shortpass dichroic filters. The reflecting cylinder and R1408 PMT are not shown in this setup. The distances and size of the various important components is shown in Figure \ref{fig:setup-schematic}.}
    \label{fig:darkbox-setup}
\end{figure}

\begin{figure}[t!]
    \includegraphics[width=0.45\textwidth]{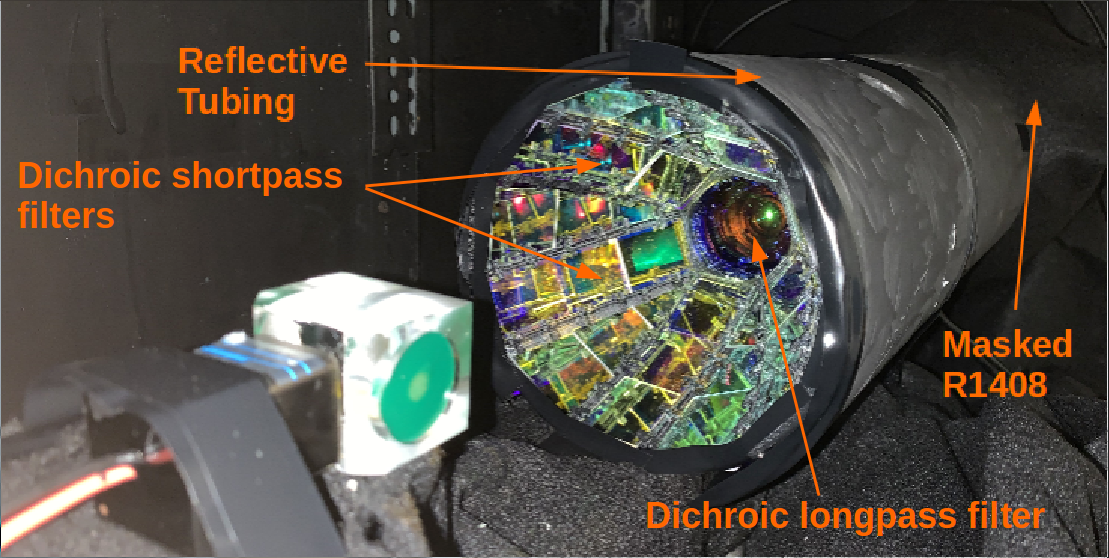}
    \caption{The full setup with the short-wavelength light detection system, which consists of a cylinder with a Mylar-lined reflecting interior that ends at an R1408 PMT. The part of the R1408 PMT outside the cylinder is masked off using black felt.}
    \label{fig:darkbox-setup2}
\end{figure}

The full setup is shown schematically in Figure \ref{fig:setup-schematic}. The front face of the R2257 or R7600-U20 is placed 215~mm away from the light source. In the complete configuration, a cylinder with reflective Mylar coating is used to direct the short-wavelength light back to an R1408 PMT, operated at 2000~V with a gain of $10^{7}$. Additionally, the R2257 or R7600-U20 are wrapped in reflective foil to ensure photons are not lost when they hit the back of central PMT and its base. The reflective cylinder is about 152~mm in diameter, to fit tightly around the dichroicon and any small gaps were closed with black tape. The R1408 PMT is an 8-inch (203~mm) diameter Hamamatsu PMT, so the outside of the PMT is masked off using felt to ensure only the central area, viewing inside the reflective cylinder, is used. Various pictures of the setup can be found in Figures \ref{fig:darkbox-setup} and \ref{fig:darkbox-setup2}. 

As discussed, this particular design for the detection of the scintillation light is not expected to be optimal. The length of the R2257 requires a long lightguide, and the R1408 PMT is larger than necessary for our narrow-view dichroicon. Our design is directed only toward achieving our primary goal here: demonstrating the sorting of photons in a way that preserves as much of both wavelength bands as possible. A more robust and integrated design would be needed for a realistic large-scale detector.

\subsection{DAQ and Data Analysis}\label{sec:data-analysis-2}

The same DAQ system as the one described in Section \ref{sec:data-analysis-1} is used. The rate of coincidences where both the trigger and aperture PMTs detect light is kept low, and thus two million triggered events were recorded for all data sets to maintain reasonable statistics. This low coincidence rate (about 1\% for the Cherenkov source) ensures that the detected events at the aperture PMT are single photoelectron (SPE). This simplifies calculating the photon arrival times where no correction for multiple photon hits is needed.

\texttt{C++}-based analysis code runs over the \texttt{hdf5} files to identify interesting events, in which light is detected by the PMTs around the dichroicon. A software-based constant fraction discriminator is used to find the time difference between the trigger PMT and the dichroicon PMTs. This is done by scanning the digitized waveform of the dichroicon PMTs looking for a threshold crossing where the voltage is three times larger than the width of the electronics noise. After each threshold crossing the number of consecutive samples above threshold are counted in a 15~ns window. If the waveform stays above threshold for longer than 3 ns, the analysis flags the threshold crossing as associated with a true PMT pulse, rather than a spike in the electronics noise. The peak of the PMT pulse is identified, and the sample associated with the 20\% peak-height crossing is found. The time of the trigger PMT is identified similarly and the large signal at the PMT makes the threshold crossing easy to find. 

In general, the dark-rate of each PMT is estimated by looking for PMT pulses in a window before the prompt light, and in all cases a correction is applied when making quantitative comparisons. This turns out to be a small correction given the relatively low dark-rates of these PMTs. 

\subsection{Dichroicon Simulation Models}\label{sec:simulations}

A simulation of the bench-top setup with the dichroicon was developed in the \texttt{Chroma} software package \cite{chroma}. \texttt{Chroma} is open-source and can be found on Github \cite{chroma-github}. \texttt{Chroma} provides a fast real-time ray-tracer, as well as a full optical Monte Carlo that can be nearly 400 times faster than \texttt{GEANT4}. The detector geometries are defined by triangulated surfaces rather than constructive solid geometry. This provides a reasonable alternative to the standard \texttt{GEANT4}-based Monte Carlo software, particularly for very large geometries such as Theia and Hyper-Kamiokande with tens of thousands of PMTs. A model of the dichroicon and bench-top setup is implement in \texttt{Chroma} and compared to the Cherenkov source data in Section \ref{sec:results-cherenkov}. Discussion of simulations of large-scale detectors with dichroicons can be found in Section \ref{sec:large-scale-simulation}.

\subsubsection{Chroma}

\begin{figure}[t!]
    \includegraphics[width=0.45\textwidth]{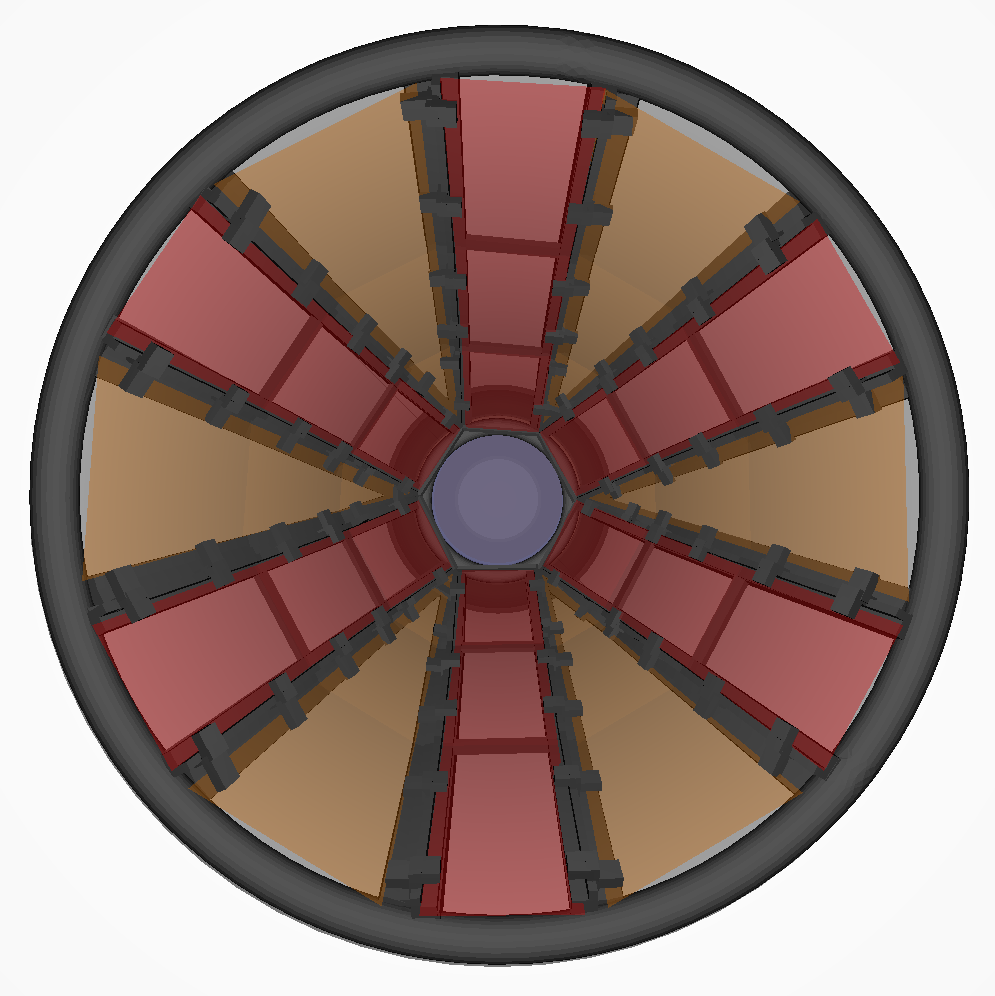}
    \caption{A direct view of the \texttt{Chroma} dichroicon model with the R2257 PMT at the center of the dichroicon. The outer diameter of the dichroicon is about 150~mm and the inner diameter, where the long-pass filter is located, is about 50~mm. The two different colors in the barrel of the dichroicon indicate the two different types of short-pass filters, as detailed in Table \ref{tab:dichroicon1}.}
    \label{fig:chroma-model}
\end{figure}

\begin{figure}[b!]
    \includegraphics[width=0.45\textwidth]{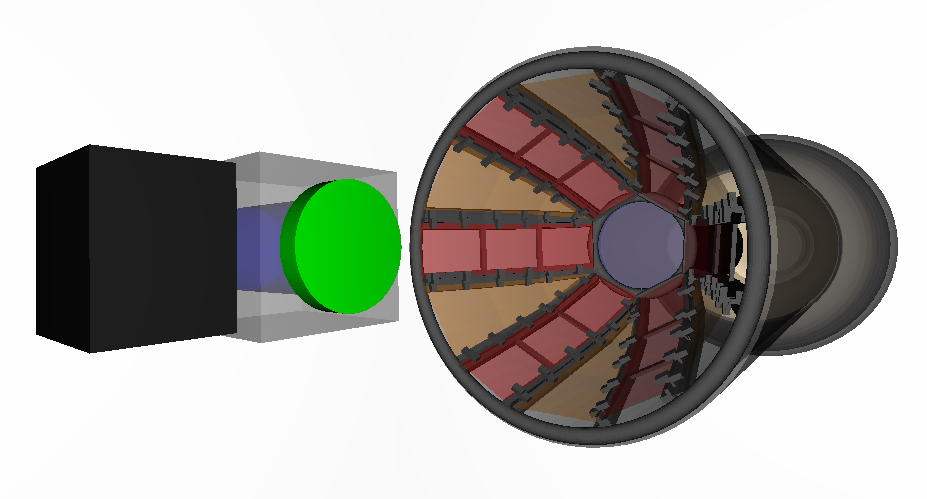}
    \caption{The full \texttt{Chroma} simulation setup for reproducing the Cherenkov source results. Geometry components with the same optical properties are colored similarly, however the colors are arbitrary. The sizes of the components and distances between the objects are identical to those used for the data, shown in Figure \ref{fig:setup-schematic}.}
    \label{fig:chroma-benchtop-model}
\end{figure}

A detailed optical model of the dichroicon is implemented in \texttt{Chroma} and shown in Figures \ref{fig:chroma-model} and \ref{fig:chroma-benchtop-model}. \texttt{Chroma} allows the triangular mesh defined by CAD software to be directly used in the simulation, so the CAD drawing for the dichroic filter holder is used to accurately reproduce the positions and orientations of the dichroic filters. For other geometry components, a triangular mesh is constructed at runtime according to measurements taken of the dichroicon. The simulation defines the surface properties of each triangle in the mesh along with the bulk properties of the material between triangles. Photons are initially produced in a \texttt{GEANT4} simulation and transferred to a GPU where they are propagated from triangle to triangle. This propagation is done on a GPU using \texttt{CUDA} ray tracing code where each \texttt{CUDA} core handles a single photon, allowing many photons to be propagated in parallel much faster than a single thread could achieve. 

\texttt{Chroma} implements several surface models that govern the behavior of photons on the triangular mesh. By default, the Fresnel equations are used to refract or reflect photons between materials of different refractive indices. A surface model that defines absorption, diffuse reflection, and specular reflection probabilities is used for the surfaces of opaque materials. Finally, a model that uses wavelength and angle-of-incidence dependent reflection and transmission probabilities is used to model the behavior of the dichroic filters. Between mesh triangles, Rayleigh scatter, attenuation, and reemission of photons is simulated according to the defined bulk material properties.

The measured transmission properties of the various filters described in Section \ref{sec:spectrometer-results} were used to create the dichroic surface models in the simulation. The filter holder mesh was set to perfectly absorbing. For the PMTs, Hamamatsu specifications were used to set the QE of the photocathodes and construct the overall geometry. A simple model of the acrylic Cherenkov source implemented in \texttt{GEANT4} to generate Cherenkov photons from the energetic electrons in Y$^{90}$ decays, which are propagated in the \texttt{Chroma} geometry.

After photons are propagated and absorbed on the photocathode, a simple DAQ simulation is performed whereby the hit times of the photons are smeared by a Gaussian distribution with a width matching the Hamamatsu provided TTS of the PMT. The earliest hit time, if any photons were detected, is taken as the hit time for the PMT in that event. Analyzing the time differences between the hit time on the tag PMT coupled to the Cherenkov source and the PMTs in the dichroicon can proceed in the same way as the measured data.

\subsection{Cherenkov Source Results}\label{sec:results-cherenkov}

The primary goal in the measurements of the Cherenkov source is to determine the effectiveness of the dichroicon for spectral sorting of Cherenkov light and, to understand how each component of the dichroicon and reflecting cylinder effects the overall performance of the full device, and to develop and test the \texttt{Chroma} model of the dichroicon. The $\Delta$t profiles presented are generated using the analysis described in Section \ref{sec:data-analysis-2}. 

Measurements are done in several staged configurations to understand the device under a variety of conditions. First, data was taken with the aperture PMT placed 215~mm away from the source, with no filters and no reflecting cylinder. This gives a baseline measurement for those to follow and is referred to as the `no filters' dataset. Second, a 480~nm dichroic longpass filter was coupled to the front-face of the aperture PMT, which acts as it would at the center of the dichroicon. In this configuration, only the dichroic longpass filter is present, and the `barrel' of the dichroicon is not included. Third, a `standard' absorbing longpass filter is added behind the dichroic longpass filter. This filter is included to absorb possible short-wavelength leakage through the dichroic filter. Finally, the barrel of the dichroicon filled with the shortpass filters was deployed. Data was taken with the dichroicon both with and without the absorbing longpass filter behind the dichroic longpass filter.

In order to compare across configurations and aperture PMTs, a factor $C^{*}_{\rm NORM}$ is defined in Equation \ref{eq:cnorm-cherenkov} that integrates the histograms, $H$, 5$\sigma$ around the mean, $\mu$, of the distributions, where $\sigma$ and $\mu$ come from a Gaussian fit to the no filters data.

\begin{equation}\label{eq:cnorm-cherenkov}
    C_{\rm NORM} = \frac{\int_{\mu - 5\sigma}^{\mu + 5\sigma} H dt}{N \times A \times \rm R_{\rm CEF}}
\end{equation}

The integral is then normalized by the number of triggered events, $N$, the photocathode area of the PMT, $A$, taken from the Hamamatsu data sheets \cite{r2257,r7600u20}, and the relative efficiencies of the PMT, R$_{\rm CEF}$. The R$_{\rm CEF}$ factor contains the collection and front-end efficiencies, which are measured relative to the R2257 PMT, as presented in Section \ref{sec:pmtcal}. $C_{\rm NORM}$ is used to compare the amount of Cherenkov light detected across different configurations and between the two aperture PMTs. By including the photocathode area, the collection efficiency, and the front-end efficiency in the $C_{NORM}$ factor, the difference in performance between the two PMTs comes primary from the different QE curves.

\begin{figure}[b!]
    \centering
    \includegraphics[width=0.45\textwidth]{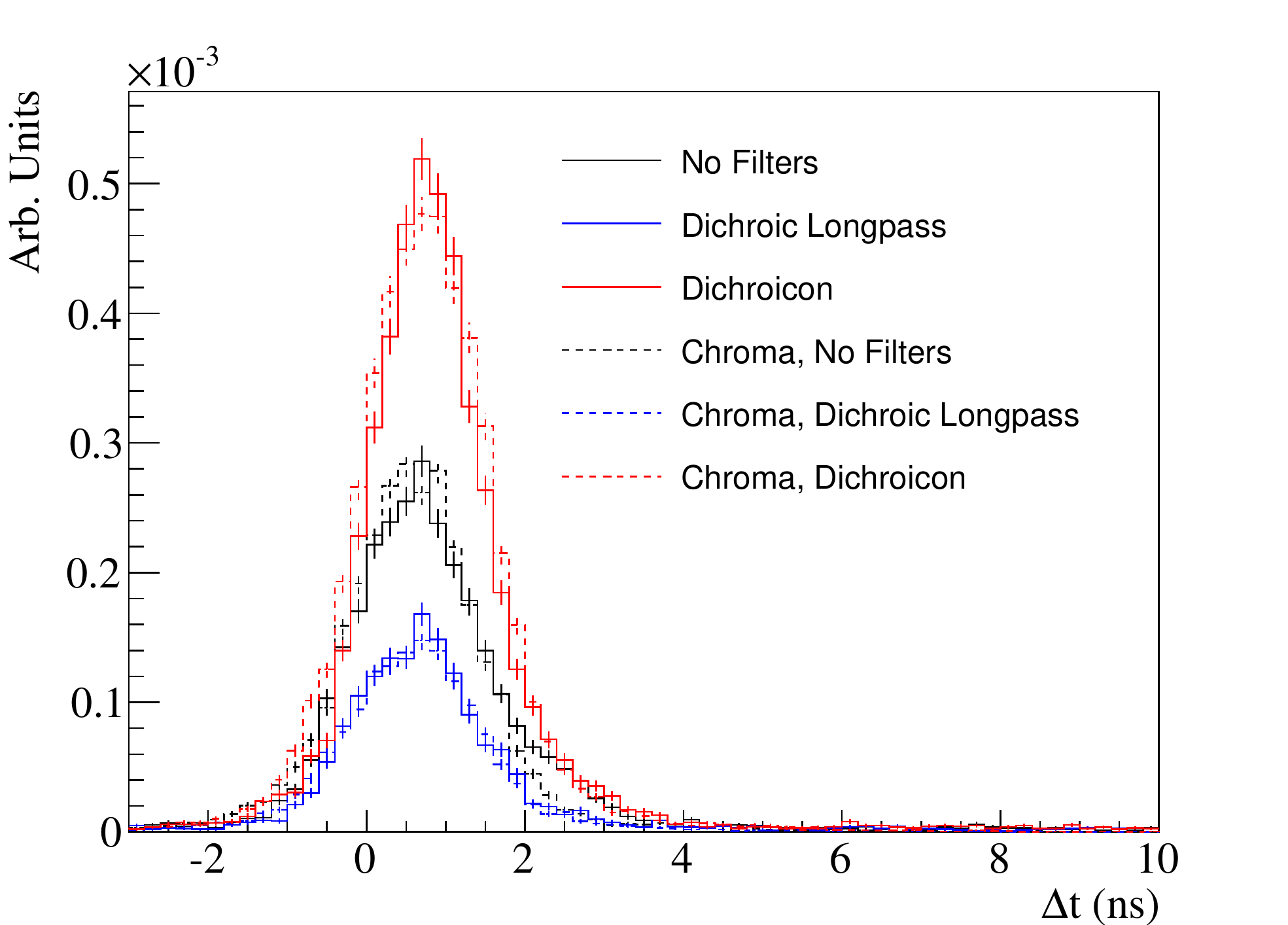}
    \caption{The results for the R2257 central aperture PMT and the acrylic Cherenkov source. In black is data for the configuration with no filters or dichroicon. The blue shows the data with the longpass dichroic filter optically coupled to the R2257. The data with the full dichroicon added is shown in red. The corresponding \texttt{Chroma} results are shown in the dashed lines. These results are summarized in Table \ref{tab:cherenkov-aperture}. The value of $\Delta$t is determined by cable delays and transit times through the PMTs and has no impact on the analysis. The results for the R7600-U20 look similar, but with a narrower TTS.}
    \label{fig:ch-results}
\end{figure}

The results of these tests for the aperture PMT are presented in Table \ref{tab:cherenkov-aperture}. In order to directly compare the \texttt{Chroma} results to the data, a scaling factor in the simulation that is used to adjust the overall efficiency is tuned so that the simulation and data agree exactly for the no filters configuration. This scaling factor is kept constant for the subsequent results. As is clear from Table \ref{tab:cherenkov-aperture} the overall agreement between data and simulation is quite good. The largest discrepancies occur for data with the dichroicon and absorbing filter, likely due small mis-modelling of the very complicated nature of the three different types of dichroic filters. Figure \ref{fig:ch-results} shows the results for the R2257 PMT under several of the configurations compared directly to the \texttt{Chroma} results. 

The results presented in Table \ref{tab:cherenkov-aperture} indicate that regardless of the configuration, the the R7600-U20 PMT detects more Cherenkov light per photocathode area than the R2257 PMT, as one would expect based on the quantum efficiencies of the PMTs. The relative changes to $C_{\rm NORM}$ depend on the shape of the QE curve of each PMT, but the qualitative features are consistent in both setups. 

\begin{table*}[t!]
    \caption{The Cherenkov source results for the R2257 and R7600-U20 aperture PMTs, with different configurations of the filters. $C_{\rm NORM}$ is defined in Equation \ref{eq:cnorm-cherenkov}. The errors are statistical only. The \checkmark indicates whether a given part of the setup was used. The first column corresponds to the central longpass dichroic filter, the second column to the absorbing longpass filter behind the dichroic filter, and the third column to the barrel of the dichroicon equipped with the shortpass dichroic filters. The results from data and the Chroma simulation are shown in the next two columns. To account for unmodeled inefficiencies in the simulation, the results for each PMT are scaled such that the case with no filters has the same $C_{\rm NORM}$ as data.}
    \label{tab:cherenkov-aperture}
    \centering 
    \begin{tabular}{lccccc}
        \hline\hline\noalign{\smallskip}
        PMT & Dichroic & Absorbing & Dichroicon & $C_{\rm NORM}$ (1/m$^{2}$) &  $C_{\rm NORM}$ (1/m$^{2}$)  \\
        & & & & [Data] & [\texttt{Chroma}]  \\
        \noalign{\smallskip}\hline\noalign{\smallskip}
        R2257 & - & - & - & 1.73 $\pm$ 0.03 & $1.73 \pm 0.02$ \\ 
        R2257 & \checkmark & - & - & 0.95 $\pm$ 0.02 & $0.91 \pm 0.01$ \\ 
        R2257 & \checkmark & \checkmark & - & 0.87 $\pm$ 0.02 & $0.78 \pm 0.01$ \\ 
        R2257 & \checkmark & - & \checkmark & 2.75 $\pm$ 0.03 & $2.94 \pm 0.03$ \\  
        R2257 & \checkmark & \checkmark & \checkmark & 1.80 $\pm$ 0.03 &  $1.65 \pm 0.02$ \\
        \noalign{\smallskip}\hline\noalign{\smallskip}
        R7600-U20 & - & - & - & 4.18 $\pm$ 0.06 &  $4.18 \pm 0.09$ \\  
        R7600-U20 & \checkmark & - & - & 1.76 $\pm$ 0.04 & $1.85 \pm 0.06$ \\ 
        R7600-U20 & \checkmark & \checkmark & - & 1.60 $\pm$ 0.04 & $1.64 \pm 0.05$ \\
        R7600-U20 & \checkmark & - & \checkmark & 8.90 $\pm$ 0.09 & $7.31 \pm 0.11$ \\  
        R7600-U20 & \checkmark & \checkmark & \checkmark & 5.80 $\pm$ 0.07 & $3.74 \pm 0.08$  \\ 
        \noalign{\smallskip}\hline\hline
    \end{tabular}
\end{table*}

Adding the dichroic longpass filter in front of these aperture PMTs reduces the Cherenkov light by about 50\%, as expected given the reflection of the shorter-wavelength photons. The reduction is slightly larger for the R7600-U20 because it has a higher sensitivity to short-wavelength light, which is largely being filtered out. Introducing the absorbing longpass filter behind the dichroic filter, an additional 10\% of the Cherenkov light is lost. This filter absorbs short-wavelength leakage through the dichroic filter, which is not designed to have perfect blocking below the cut-on wavelength, particularly for photons at large incidence angles. This becomes important for the measurements presented in Section \ref{sec:results-labppo}, where the absorbing filter is used to remove scintillation light leakage through the dichroic filter.

By incorporating the dichroicon barrel, filled with shortpass dichroic filters, the total amount of Cherenkov light collected at the aperture (through only the dichroic longpass filter) is increased by a factor of 2.9 and 5.1 for the R2257 and R7600-U20 respectively. The larger factor for the R7600-U20 comes from the relatively larger short-wavelength sensitivity---the short-wavelength light, which can be reflected by the barrel of the dichroicon, now strikes the central dichroic filter at higher incidence angles, where the filter is more likely to leak the photon. 

In addition to detecting the light at the aperture, as with a standard Winston cone, our design requires the detection of the light transmitted through the short-pass dichroic filters. The short-wavelength system, consisting of the reflective cylinder and R1408 PMT, was tested with both aperture PMTs. The inclusion of this system did not affect the amount of light detected by either aperture PMT. The R1408 detects 3.3 times more light than either aperture PMT, after correcting for the measured relative differences in the front-end and collection efficiencies. 

Simulations of this setup over-predict the total light detected by the R1408 by a factor of 3. This is likely due to the fact that the model does not include many inefficiencies in the setup of the short-wavelength detection system. These primarily include non-perfect reflectivity of many of the components, non-uniform deployment of the reflective components, and the lack of the inclusion of the PMT stands in the model. The result is nevertheless encouraging, as this simple system detects 33\% of the predicted short-wavelength light. A more integrated device with better and well-understood reflective coating could easily improve on these results and should in principle be easier to model. 

Overall, with both aperture PMTs, the dichroicon demonstrated excellent capabilities for spectral sorting of photons towards two different PMTs, simultaneously detecting both the short- and long-wavelength light.

\subsection{LAB+PPO Results}\label{sec:results-labppo}

The results presented in this section use the setup described in Section \ref{sec:experimentalsetup}, now with the LAB+PPO scintillation source. As discussed, one of the primary goals with this source will be to separate the scintillation and Cherenkov components of the light emission. Figure \ref{fig:wavelengths}, which compares the shape of the PPO emission spectrum to the shape of the Cherenkov emission spectrum, illustrates that photons with wavelengths above 500~nm should be primarily Cherenkov light. In the full setup, this light will be directed toward the aperture PMT, while the shorter wavelength scintillation light will pass through the dichroicon and be detected by the R1408. As with the Cherenkov measurements, presented in Section \ref{sec:results-cherenkov}, measurements are made with the two different aperture PMTs.

The $\Delta t$ distributions are created using the analysis techniques described in Section \ref{sec:data-analysis-2}. The scintillation time profile for LAB+PPO is well understood and has been measured in \cite{OKeeffe:2011dex,Lombardi:2013nla,MarrodanUndagoitia:2009kq}, which we use as a guide for our fits to the time profiles. As discussed in \cite{OKeeffe:2011dex}, deoxygenation of the LAB+PPO leads to a reduction in the quenching of the scintillation light, which primarily affects the late-light timing in the tail of the $\Delta$t distribution. The focus of these measurements is primarily on the prompt light and we fit our spectrum only to 20~ns past the prompt peak, so deoxygenation of the scintillator is not performed. 

Following the procedure with the Cherenkov source, we first make measurements of light at the aperture PMT under varying conditions. These includes incrementally adding an absorbing longpass filter and then the barrel of the full dichroicon, keeping track of the relative effect of each component. Unlike the Cherenkov source, however, we cannot just integrate the prompt peak, because we may have significant contamination from scintillation light leaking through the dichroic filter. Additionally, unlike the Cherenkov source measurement, no measurement is made without the central dichroic filter, as the Cherenkov light is completely dominated by the large scintillation light yield.

The $\Delta$t spectra for the central PMT are fit using Equation \ref{eq:fit} using the \texttt{RooFit} package \cite{Verkerke:2003ir}.

\begin{multline}\label{eq:fit}
F =  C \times f_{PMT}(t - t') + \\ (1 - C) \times \sum_{i=1}^{2} \frac{A_{i} \times (e^{-t/\tau_{i}} - e^{-t/\tau_{R}})}{(\tau_{i} - \tau_{R})} * f_{PMT}(t - t')
\end{multline}

The fit is performed to the prompt Cherenkov and scintillation light between 0 to 30~ns in the $\Delta$t histogram. The scintillation component of the fit uses the sum of two decay exponentials with time constants $\tau_{1}$ and $\tau_{2}$ and associated weights $A_{1}$ and $A_{2}$ and an exponential rise time $\tau_{R}$. The weights are constrained to sum to one. The PMT TTS, $f_{PMT}$, is determined from the Cherenkov source data to be around 350~ps for the R7600-U20 and 900~ps for the R2257. The mean, $\mu$, and width, $\sigma$, of $f_{PMT}$ are constrained to the values found when performing Gaussian fits to the Cherenkov source data. This distribution is convolved with the scintillation time-profile in the fit. There is an arbitrary offset, $t'$, which allows for cable delays and other time-offsets, that is different for the two PMTs. The Cherenkov component is modeled by weighting the PMT TTS distribution by an appropriate factor $C$. The Cherenkov and scintillation components are constrained to sum to one. 

\begin{figure}[t!]
    \centering
    \includegraphics[width=0.45\textwidth]{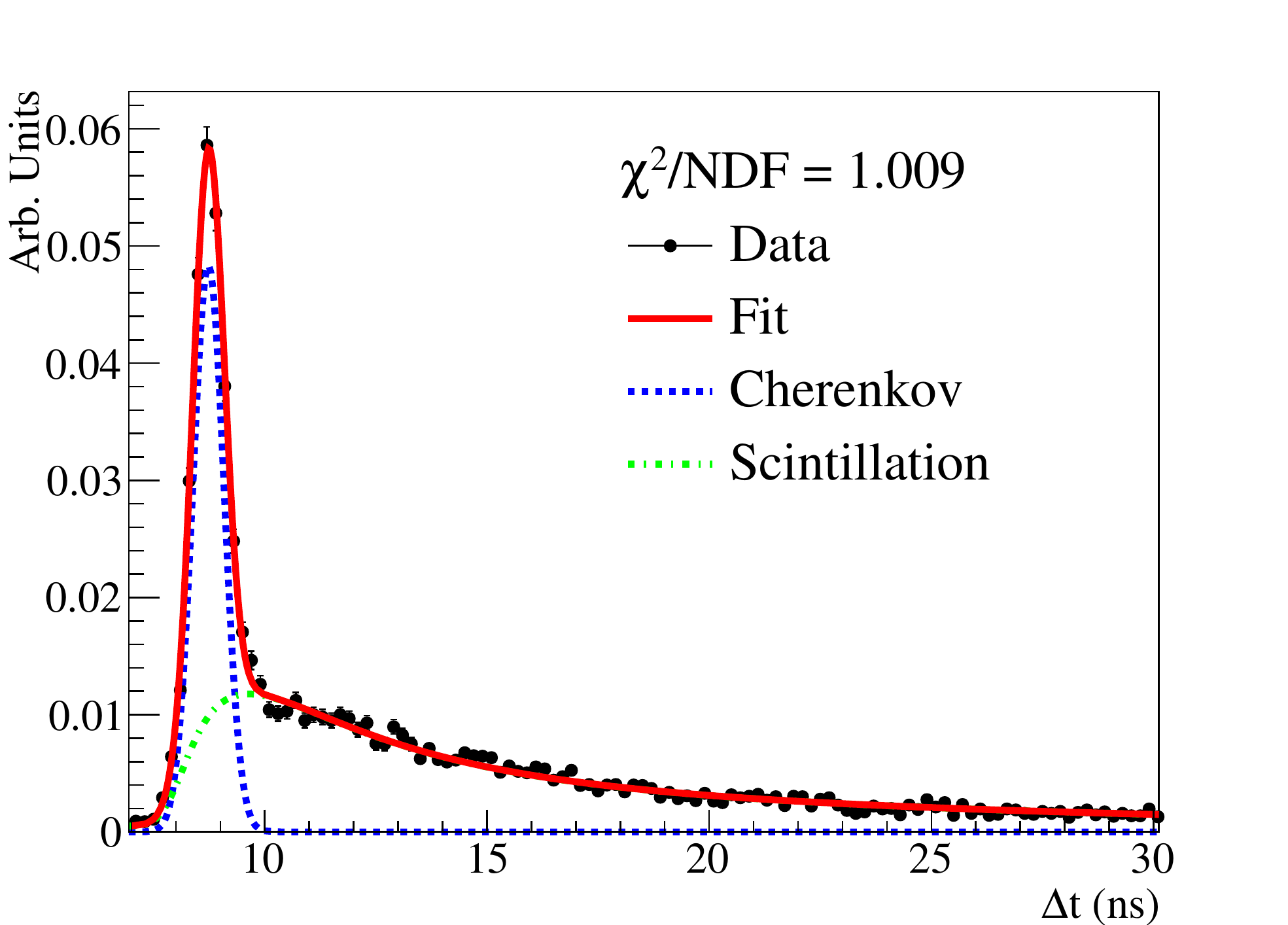}
    \caption{An example fit using Equation \ref{eq:fit}, to LAB-PPO data. The Cherenkov and scintillation components are shown separately, in addition to the total fit in red. The data were taken with the shortpass barrel of the dichroicon and the absorbing longpass filter behind the dichroic longpass filter at the aperture of the dichroicon. The data is normalized to 1.0. The $\Delta$t offset from zero is arbitrary and does not impact the fit.}
    \label{fig:example-fits}
\end{figure}

Figure \ref{fig:example-fits} shows an example fit done to the LAB+PPO data with the R7600-U20 at the aperture of the dichroicon, with the Cherenkov and scintillation components shown separately. As is evident, the scintillation and Cherenkov components of the light are very nicely separated, and a prompt timing window can be selected to identify a pure sample of Cherenkov light. The purity of the selection of the prompt Cherenkov light, $P$, is defined in Equation \ref{eq:purity}, which calculates the fraction of the Cherenkov component of the fit in a non-symmetric window that goes from $-5\sigma$ to $+ 1.5\sigma$ around the mean of the distribution.

\begin{equation}\label{eq:purity}
P = \int_{\mu - 5\sigma}^{\mu + 1.5\sigma} \frac{C \times f_{PMT}(t - t')}{F} dt
\end{equation}

In addition to the purity, the total amount of Cherenkov light detected is an interesting quantity that can be directly compared to the Cherenkov source data. Given that the indices of refraction of LAB+PPO and acrylic are almost identical \cite{Yeo:2010zz}, the total amount of Cherenkov light collected should be nearly the same. As with the Cherenkov source data, we correct for the number of triggered events, the photocathode area of the PMT, and the collection and front-end efficiencies of the PMT, as defined in Equation \ref{eq:cnorm}. Note this equation is essentially identical to Equation \ref{eq:cnorm-cherenkov}, but here we are explicitly using the fit to remove the expected counts from scintillation leakage into the prompt window. 

\begin{equation}\label{eq:cnorm}
    C^{*}_{\rm NORM} = \frac{\int_{\mu - 5\sigma}^{\mu + 5\sigma} C \times f_{PMT}(t - t') dt}{\rm N \times \rm A \times \rm R_{\rm CEF}}
\end{equation}

Particle identification and reconstruction in a real scintillation detector using Cherenkov light will depend on both the purity of the Cherenkov photons and their total number. One can achieve great purity by, for example, moving the cut-on wavelength toward even longer wavelengths, but this buries more of the remaining Cherenkov photons beneath the scintillation light and thus would lead to only a small number of usable Cherenkov photons.  Thus we use as one metric the total number of Cherenkov photons multiplied by the purity, 
\begin{equation}\label{eq:ratios}
    R = C^{*}_{\rm NORM} \times P.
\end{equation}

When $P=1$ the detector can operate identically to a Cherenkov detector, up to the total light yield given by $C^{*}_{\rm NORM}$. In a real detector, additional information can be used to identify Cherenkov photons (such as an hypothesized event direction) and thus while $R$ provides a good relative metric, it may be a pessimistic assessment of the total effective Cherenkov yield.

The results for both aperture PMTs for the various configurations are presented in Table \ref{tab:central-fits} with the corresponding \texttt{Chroma} results in Table \ref{tab:central-fits-chroma}. The $C^{*}_{\rm NORM}$ results agree well with the Cherenkov source data and can be directly compared to the $C_{\rm NORM}$ values in Table \ref{tab:cherenkov-aperture}. As with the Cherenkov source results, the \texttt{Chroma} predictions agree well with the data, particularly for the R2257. Again, the largest discrepancy is with the dichroicon with the R7600-U20, where the total amount of Cherenkov light collected is being under predicted. Additionally, the \texttt{Chroma} results systematically under predicts the amount of scintillation light leakage through the central dichroic longpass filter, leading to a higher purity selection of the Cherenkov light in the prompt window. Future efforts to characterize the dichroic filters in more detail should help improve the agreement between the data and simulation.

\begin{table*}[t!]
    \caption{Results for the LAB+PPO source with the R2257 and R7600-U20 central PMTs. $C^{*}_{\rm NORM}$ is total Cherenkov light, normalized to the number of triggers and photocathode area, defined in Equation \ref{eq:cnorm}. $P$ is the purity of the Cherenkov light selection in a prompt window, defined in Equation \ref{eq:purity}. $R$ is $C^{*}_{\rm NORM} \times P$. The errors come from the uncertainties on the fit parameters. The \checkmark indicates whether a given part of the setup was used. The first column corresponds to the central longpass dichroic filter, the second column to the absorbing longpass filter behind the dichroic filter, and the third column to the barrel of the dichroicon equipped with the shortpass filters. } 
    \label{tab:central-fits}
    \centering 
    \begin{tabular}{lcccccc}
         \hline\hline\noalign{\smallskip}
         PMT & Dichroic & Absorbing & Dichroicon & $C^{*}_{\rm NORM}$ (1/m$^{2}$) & $P$ (\%) & $R$ \\ 
         \noalign{\smallskip}\hline\noalign{\smallskip}
         R2257 & \checkmark & - & - & 0.91 $\pm$ 0.06 & 70.8 $\pm$ 1.4 & 0.64 $\pm$ 0.06 \\
         R2257 & \checkmark & \checkmark & - & 0.80 $\pm$ 0.06 & 90.6 $\pm$ 1.1 & 0.72 $\pm$  0.06 \\
         R2257 & \checkmark & - & \checkmark  & 2.77 $\pm$ 0.13 & 82.2 $\pm$ 1.2 & 2.28 $\pm$ 0.07 \\
         R2257 & \checkmark & \checkmark & \checkmark  & 1.84 $\pm$ 0.10 & 90.0 $\pm$ 1.1 & 1.66 $\pm$ 0.07 \\
         \noalign{\smallskip}\hline\noalign{\smallskip}
         R7600-U20 & \checkmark & - & - &  1.77 $\pm$ 0.13 & 69.9 $\pm$ 1.5 & 1.24 $\pm$ 0.09 \\ 
         R7600-U20 & \checkmark & \checkmark & - &  1.73 $\pm$ 0.13 & 95.5 $\pm$ 1.4 & 1.65 $\pm$ 0.13 \\
         R7600-U20 & \checkmark & - & \checkmark & 8.71 $\pm$ 0.64 & 84.4 $\pm$ 1.5 & 7.35 $\pm$ 0.56 \\
         R7600-U20 & \checkmark & \checkmark & \checkmark & 5.54 $\pm$ 0.40 & 93.2 $\pm$ 1.3 & 5.16 $\pm$ 0.38 \\
         \noalign{\smallskip}\hline\hline
    \end{tabular}
\end{table*}

\begin{table*}[t!]
    \caption{Results for the \texttt{Chroma} simulations of the LAB+PPO source with the R2257 and R7600-U20 central PMTs. $C^{*}_{\rm NORM}$ is total Cherenkov light, normalized to the number of triggers and photocathode area, defined in Equation \ref{eq:cnorm}. $P$ is the purity of the Cherenkov light selection in a prompt window, defined in Equation \ref{eq:purity}. $R$ is $C^{*}_{\rm NORM} \times P$. The errors come from the uncertainties on the fit parameters. The \checkmark indicates whether a given part of the setup was simulated. The first column corresponds to the central longpass dichroic filter, the second column to the absorbing longpass filter behind the dichroic filter, and the third column to the barrel of the dichroicon equipped with the shortpass filters. To account for unmodeled inefficiencies in the simulation, the results for each PMT are scaled such that the case with only the longpass dichroic filter has the same $C^{*}_{\rm NORM}$ as data.} 
    \label{tab:central-fits-chroma}
    \centering 
    \begin{tabular}{lcccccc}
         \hline\hline\noalign{\smallskip}
         PMT & Dichroic & Absorbing & Dichroicon & $C^{*}_{\rm NORM}$ (1/m$^{2}$) & $P$ (\%) & $R$ \\ 
         \noalign{\smallskip}\hline\noalign{\smallskip}
         R2257 & \checkmark & - & - & 0.91 $\pm$ 0.10 & 87.8 $\pm$ 1.1 & 0.80 $\pm$ 0.11 \\
         R2257 & \checkmark & \checkmark & - & 0.83 $\pm$ 0.08 & 97.7 $\pm$ 1.1 &  0.81 $\pm$ 0.08 \\
         R2257 & \checkmark & - & \checkmark  & 2.60 $\pm$ 0.21 & 86.2 $\pm$ 1.4 & 2.24 $\pm$ 0.25 \\
         R2257 & \checkmark & \checkmark & \checkmark & 1.72 $\pm$ 0.16 & 96.4 $\pm$ 1.1 & 1.66 $\pm$ 0.17 \\
         \noalign{\smallskip}\hline\noalign{\smallskip}
         R7600-U20 & \checkmark & - & - & 1.77 $\pm$ 0.22 & 80.3 $\pm$ 3.6 & 1.42 $\pm$ 0.29 \\ 
         R7600-U20 & \checkmark & \checkmark & - & 1.76 $\pm$ 0.21 & 95.0 $\pm$ 1.8 & 1.67 $\pm$ 0.22 \\
         R7600-U20 & \checkmark & - & \checkmark & 6.41 $\pm$ 0.49 & 80.0 $\pm$ 1.7 & 5.13 $\pm$ 0.64 \\
         R7600-U20 & \checkmark & \checkmark & \checkmark & 3.63 $\pm$ 0.32 & 96.4 $\pm$ 1.9 & 3.50 $\pm$ 0.34 \\
         \noalign{\smallskip}\hline\hline
    \end{tabular}
\end{table*}

From these tables it is clear that an excellent purity is achieved with both the R2257 and R7600-U20 PMTs. As expected, given the faster timing of the R7600-U20 PMT, we are able to achieve better purity with this PMT. In all setups with the absorbing longpass filter behind the longpass dichroic filter at the aperture, the purity for Cherenkov selection in the prompt window is better than 90\%. It is important that this purity is achieved with a prompt window that still selects that majority of the Cherenkov light -- that is, to achieve a higher purity of Cherenkov selection we are not selecting an extremely narrow window around the prompt peaks, which would reject not only the scintillation light, but also large amount of the Cherenkov light. 

By comparing the $R$ values for the various datasets we can make several conclusions. First, comparing the first two rows for each PMT in Table \ref{tab:central-fits}, the introduction of the 500~nm absorbing longpass filter without the barrel of the dichroicon increases the purity without significantly affecting the total Cherenkov light detected, effectively increasing $R$. However, when the full dichroicon is in place, the total Cherenkov light detected is significantly affected by the inclusion of the absorbing longpass filter, which was discussed in the Cherenkov source data and is due to the leakage of photons through the central dichroic filter at high incidence angles. This suggests that the inclusion of the absorbing longpass filter is not necessarily the optimal configuration for a large-scale detector, where not just purity but the total number of detected Cherenkov photons is important. While the purity of the selection is increased (by about 10\%) with the absorbing longpass filter, the total Cherenkov light lost is about 30\%, so the value of $R$ decreases. 

Perhaps most notably, by including the full dichroicon we improve our Cherenkov light collection by about a factor of 5 for the R7600-U20 and a factor of 3 for the R2257, identical to the Cherenkov source results. This suggests that having a red-sensitive PMT is important at the center of the dichroicon, but having additional blue sensitivity will help with the detection of the Cherenkov light. The temporal separation with these PMTs is already quite good, and therefore a small amount of short-wavelength scintillation leakage is acceptable, as both Cherenkov and scintillation light is leaked, so one increases the total Cherenkov light detected while sacrificing a small amount of purity. 

\begin{figure}[ht!]
    \centering
    \includegraphics[width=0.45\textwidth]{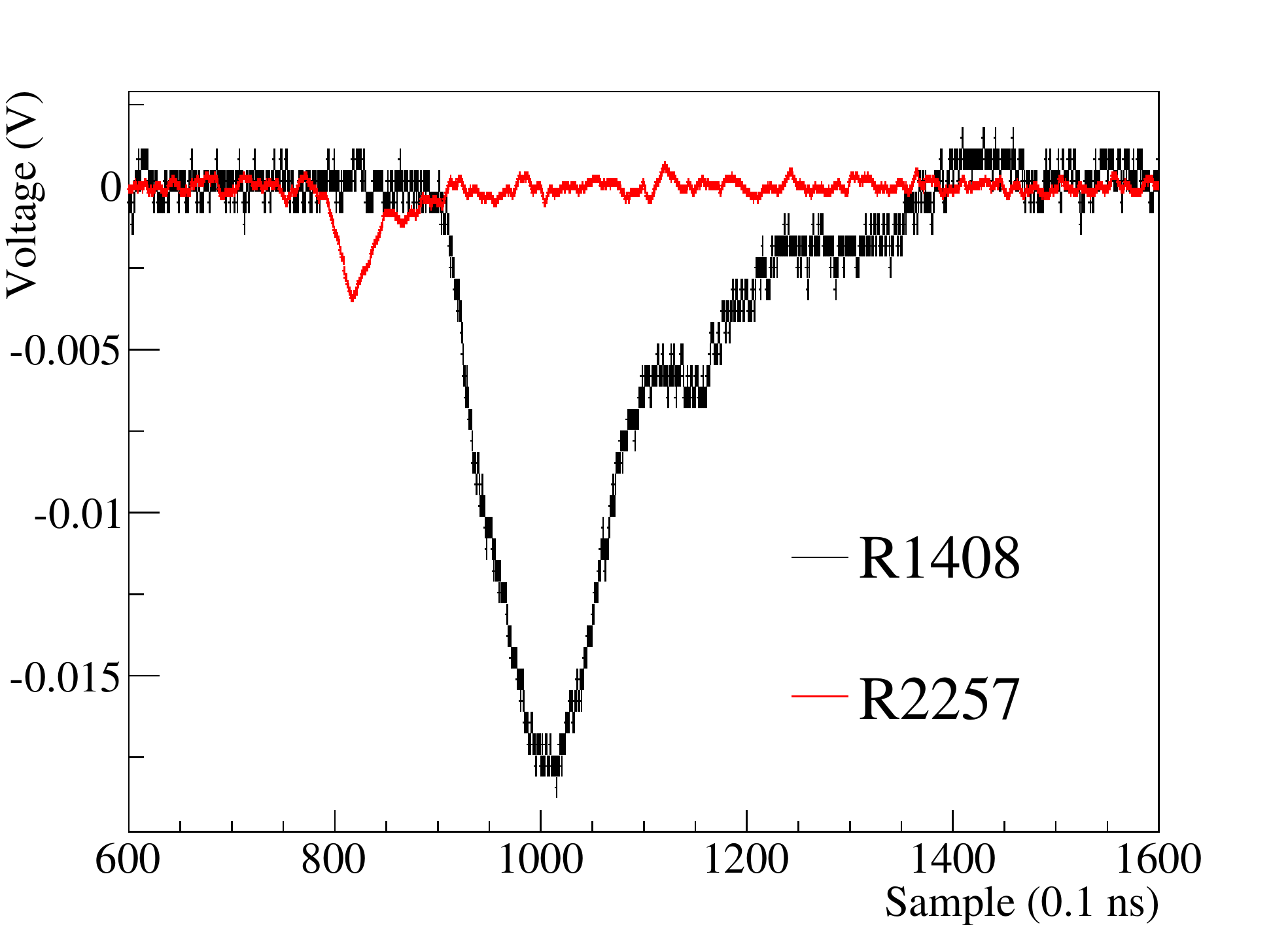}
    \caption{The R2257 and R1408 waveforms for a single triggered event. This event has an early-time PMT pulse, corresponding to Cherenkov light, for the R2257 PMT and an approximately 5 PE pulse at the R1408 PMT. The waveforms are specifically selected by looking at the $\Delta t$ histogram and selecting an event in the Cherenkov peak. The $x$-axis is labelled in 0.1~ns samples, and the offset between the two waveforms is showing the additional photon travel time and transit time through the R1408 PMT.}
    \label{fig:waveforms}
\end{figure}

In addition to the Cherenkov light detection, the ability to sort scintillation light to the back R1408 PMT is a critical feature of the dichroicon design. The setup is adjusted to include the full reflective cylinder and R1408 PMT. Encouragingly, it was easy to identify individual events with coincidences of Cherenkov light at the aperture PMT and scintillation light at the back PMT. A typical waveform for the setup with the R2257 is shown in Figure \ref{fig:waveforms}. In this event, it is clear that the back PMT detected several photons, leading to a PMT pulse size corresponding to around 5 photoelectrons, while the aperture PMT detected a single Cherenkov photon. The scintillation timing profile is also evident in the pulse shape of the R1408.

The data for the aperture PMT is compared with and without the reflecting cylinder, and no difference is found. This indicates that the reflecting light guide does not lead to additional scintillation light bouncing off of the cylinder and leaking through the central dichroic filter. The total light collected at the R1408 PMT should represent the high light yield of the scintillator. To quantify this, we integrate the charge spectrum, which is heavily multi-PE, using Equation \ref{eq:charge-tot}.

\begin{equation}\label{eq:charge-tot}
Q_{\rm TOT} = \sum_{i=0}^{N} \frac{ Q_{i} \times C_{i} }{ Q_{\rm SPE} }
\end{equation}

Here $Q_{i}$ is the charge of bin $i$ in pC and $C_{i}$ is the associated bin content. This sum is normalized by the charge of the SPE peak $Q_{\rm SPE}$, which is 1.6~pC as the PMT is operating at a gain of $10^{7}$. This is a good measure of the total light collected by the R1408 and can be compared to the total light collected by the aperture PMT, which remains in the SPE regime and the photon counting is done the same way as with the Cherenkov source. The R1408 detects about 550 times more light than both aperture PMTs in these setups. This highlights the traditional difficulty with separating the two components of the light -- the scintillation yield overwhelms the Cherenkov yield. Using the dichroicon we simultaneously detect a large fraction of the scintillation light with the R1408, while detecting the Cherenkov light at the aperture with high purity. To quantify this, we compare our results for the R1408 to the \texttt{Chroma} prediction and find an efficiency for detecting the short-wavelength light around 30\%, consistent within uncertainties to the value found for the Cherenkov source. With a better engineered, integrated system we expect to be able to increase this efficiency for detecting the short-wavelength light.

\begin{figure}[b!]
\centering 
\includegraphics[width=0.45\textwidth]{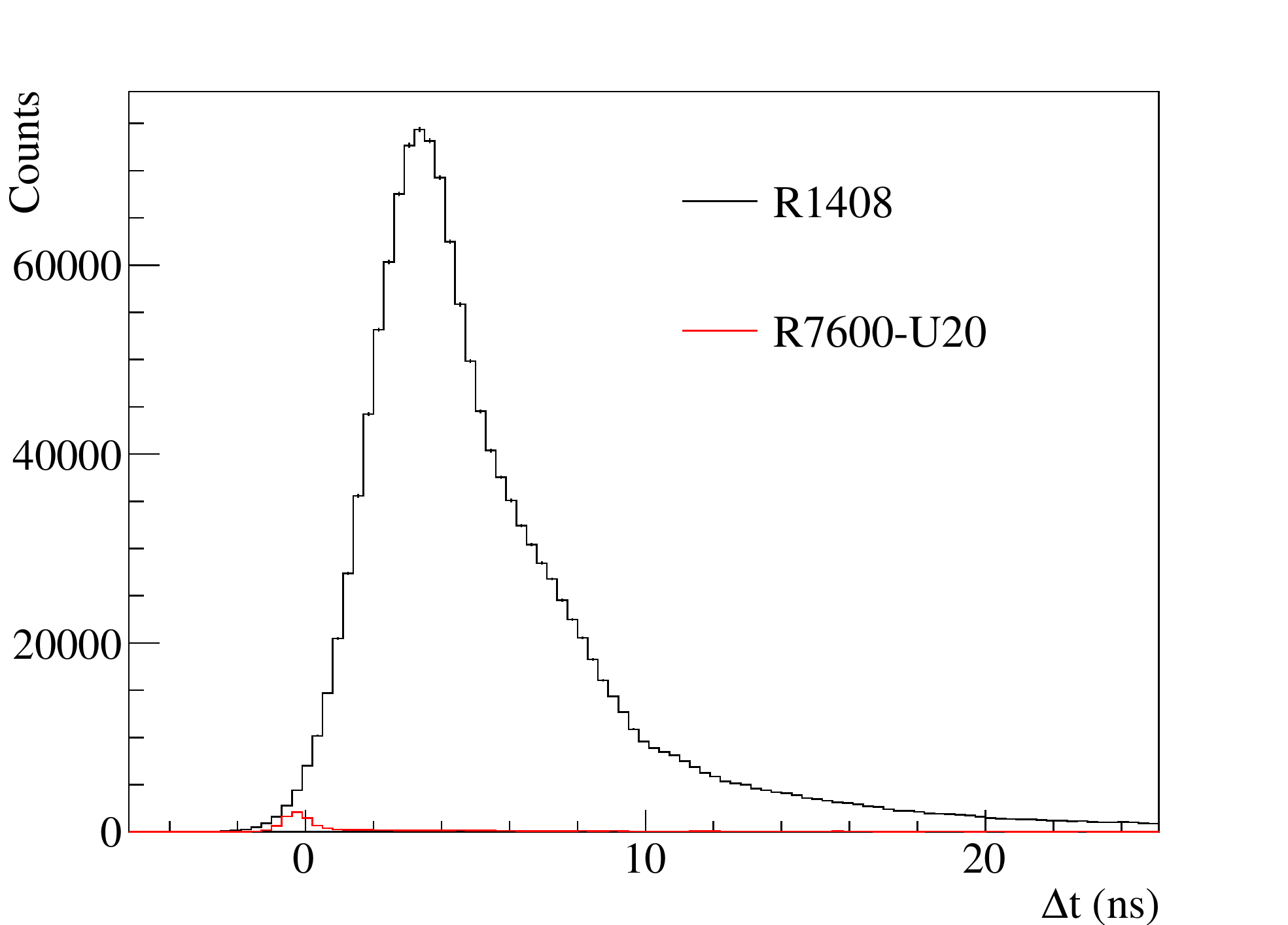}
\caption{The dichroicon data for both the R7600-U20 PMT at the aperture and the R1408 PMT behind the dichroicon. The light detected by the R1408 is primarily scintillation light and the light detected by the R7600-U20 is primarily Cherenkov light.}
\label{fig:scint-dichroicon-reaout}
\end{figure}

The time distribution of the detected light for the R7600-U20 and R1408 PMTs can be seen in Figure \ref{fig:scint-dichroicon-reaout}. The scintillation light detected by the R1408 PMT swamps the Cherenkov light collected by the R7600-U20 PMT, making it very clear why this separation is so difficult without spectral photon sorting. This figure clearly illustrates the power of the dichroicon -- by detecting the Cherenkov and scintillation light in separate PMTs we are able to maintain  high light yield of the scintillator while simultaneously detecting the Cherenkov light. 

\subsubsection{Pulse Shape Discrimination (PSD)}\label{sec:psd}

For liquid scintillator detectors, particle identification (PID) is critical for background rejection. LAB+PPO is well known to have good pulse-shape discrimination between $\beta$ and $\alpha$ excitation, a thorough discussion of which can be found in \cite{Lombardi:2013nla}. In short, $\alpha$ particles are known to excite the slow component of the scintillation emission, which is typically associated with triplet state excitation of the solute molecules, more than $\beta$ particles. The ratio of the amount of prompt to the amount of late light can be used as a handle to separate $\beta$ from $\alpha$ particles. A practical application of this PSD technique in the Borexino detector can be found in \cite{Galbiati:2016inv}, and is discussed for SNO+ and LENA in \cite{Andringa:2015tza, Wurm:2011zn}. Similar PSD techniques for background rejection in scintillators can be found in \cite{Ashenfelter:2015aaa, Amaudruz:2016qqa}.

In principle, there is another difference in the light production for $\beta$ and $\alpha$ particle excitation in liquid scintillator. Most $\beta$ particles from radioactive decays are above the Cherenkov threshold, whereas the $\alpha$ particles are not. Thus, if Cherenkov light could be identified at energies around a couple of MeV in a liquid scintillator detector, the absence of Cherenkov light would be a clear tag for $\alpha$ excitation.

We tested this in our setup using a $^{210}$Po $\alpha$ source to irradiate the LAB+PPO, replacing the $^{90}$Sr source. For this, we used only the R7600-U20 aperture PMT, as both central PMTs behave well in terms of detecting ample Cherenkov light in the setup with the dichroicon. The $^{210}$Po decays 100\% of the time via a 5.41 MeV $\alpha$, which enters the scintillator, creating scintillation light. 

\begin{figure}[t!]
    \centering
    \includegraphics[width=0.45\textwidth]{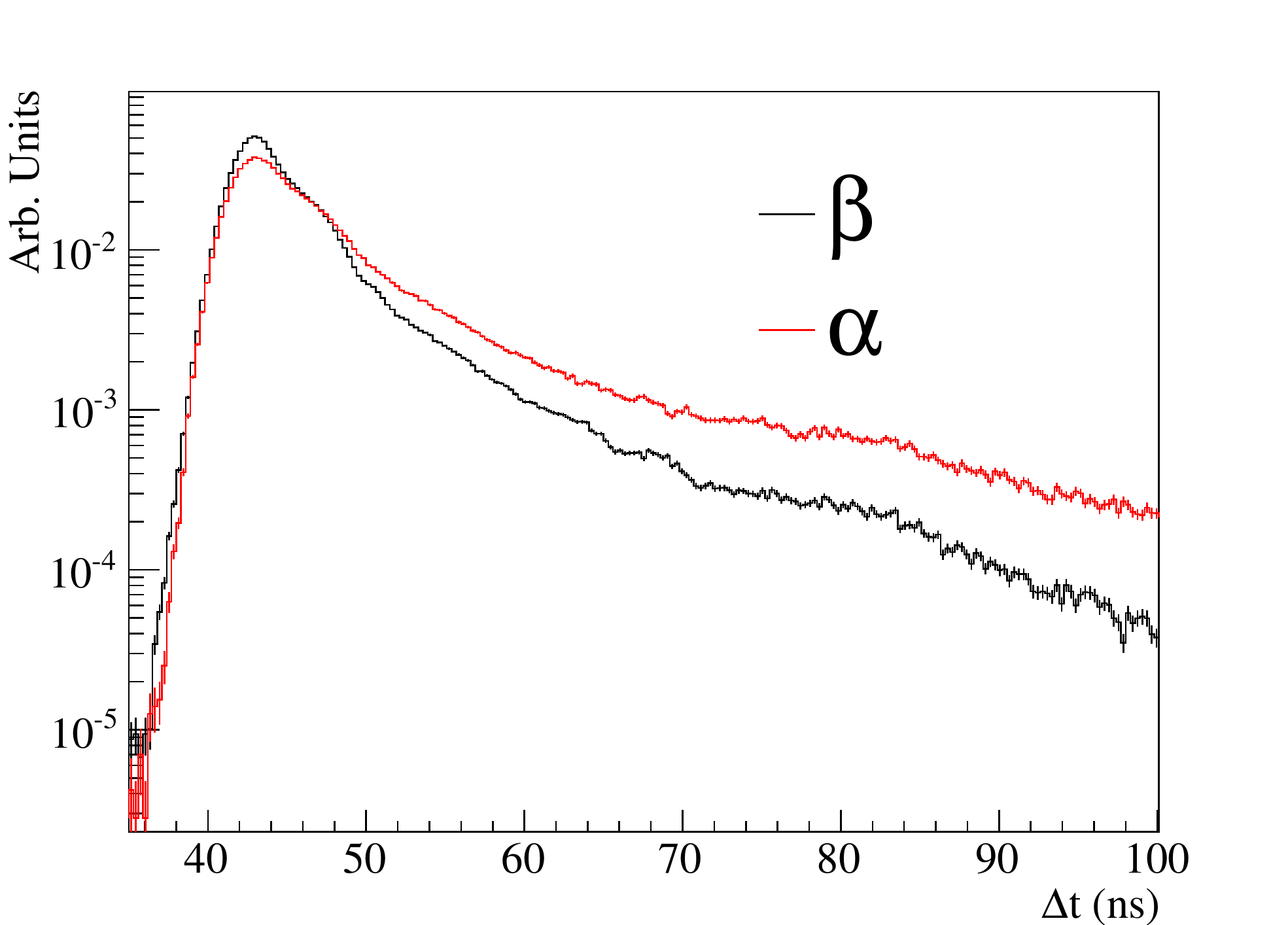}
    \caption{The R1408 data for a $^{90}$Sr $\beta$ source and a $^{210}$Po $\alpha$ source. The small bumps in the timing spectrum are due to the complicated PMT transit time distribution, which includes two different late-pulsing peaks. The difference between these distributions is used to discriminate between $\beta$ and $\alpha$ particle excitation in liquid scintillator detectors.}
    \label{fig:alpha-r1408}
\end{figure}

Figure \ref{fig:alpha-r1408} shows the data for the R1408 PMT, which detects the short-wavelength scintillation light through the barrel of the dichroicon. As expected, the typical difference in the scintillation time-profiles is identified, where, under $\alpha$ excitation, the scintillator produces more late light. This is the typical manner in which liquid scintillator detectors perform PID.

\begin{figure}[b!]
    \centering
    \includegraphics[width=0.45\textwidth]{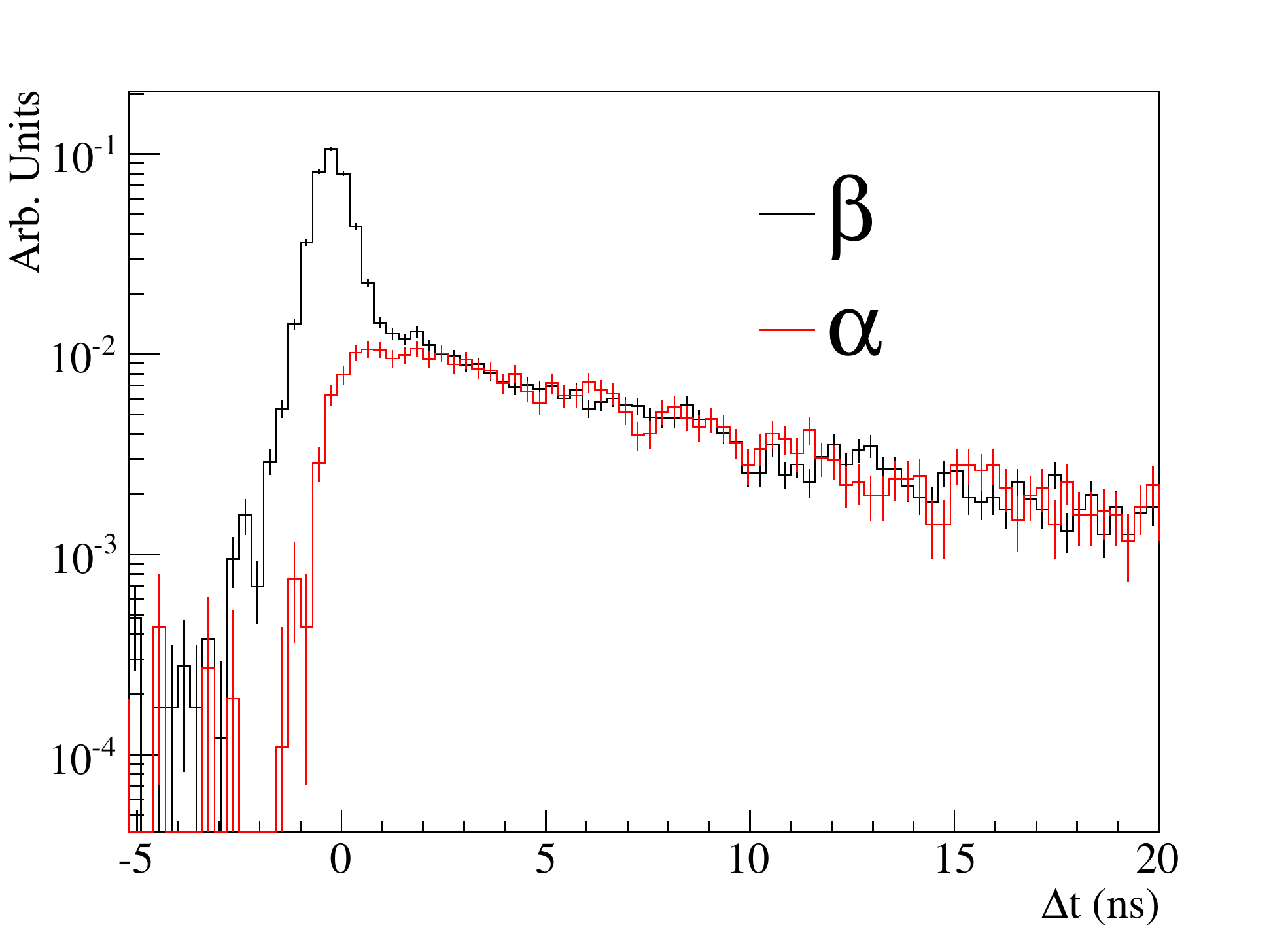}
    \caption{The R7600-U20 data for a $^{90}$Sr $\beta$ source and a $^{210}$Po $\alpha$ source. The lack of Cherenkov light for the below-threshold $\alpha$ particles can be clearly identified.}
    \label{fig:alpha}
\end{figure}

The data for the R7600-U20 at the aperture of the dichroicon is shown in Figure \ref{fig:alpha}. As can be seen clearly, the prompt Cherenkov light is absent in the data with the $\alpha$ source. Using the full dichroicon setup, both the difference in the scintillation time-profiles for the back PMT and the difference at early times for the aperture PMT can be used to discriminate between $\alpha$ and $\beta$ particles. 

\subsubsection{Off-Axis Source}\label{sec:offaxis}

Two measurements with the source displaced from center of the dichroicon are performed with the LAB+PPO scintillation source to understand the behavior of the dichroicon. In all cases the scintillation light is emitted isotropically, so as the source moves further from the central-axis of the dichroicon, the average incidence angle of the photons increase and the solid-angle acceptance shrinks. The coordinate system is defined in Figure \ref{fig:setup-schematic} and is used to describe the movement of the source. In the first off-axis test the source is relocated to the edge of dichroicon, moving it  75~mm in the +y direction. The second off-axis test moves the source 75~mm further in the +y direction. The x-distance from the dichroicon is kept constant at 115~mm and the z-position is kept at the same level as the center of the dichroicon. 

In the first off-axis measurement we find a reduction consistent with 50\% in the light collection of the Cherenkov light at the aperture PMT, as expected based on the change in solid angle. For the second off-axis measurement the source is now outside the geometric field of view, so only 6\% of the Cherenkov light is collected. The only reason we collect any light at all in this setup is that our design is not a perfect Winston cone.

For these off-axis measurements, we again quantify the total light collected at the R1408 PMT using $Q_{\rm TOT}$, as defined in Equation \ref{eq:charge-tot} and compare it relative to the on-axis data. The data collected at this PMT does not require the Winston cone to work as a reflector---instead, it only relies on the short-wavelength light incident on the dichroicon to be transmitted and successfully reflected back to the PMT. For the off-axis 1 data we again find about 50\% of the total light detected. Then for off-axis 2 data we find now 29\% of the light still collected, a much larger factor than found for the aperture PMT. This demonstrates the short-wavelength light collection is more robust to large angles of incidence, outside of the view of the Winston cone. Table \ref{tab:off-axis-results} summarizes the results presented in this section.

\begin{table}[t!]
    \caption{A summary of the results for the off-axis measurements with the R2257 PMT at the aperture of the dichroicon. The value of $C^{*}_{\rm NORM}$ for the central data is the same as given in Table \ref{tab:central-fits}. The $C^{*}_{\rm NORM}$ and $Q_{\rm TOT}$ values for the first off-axis measurement at y of 75~mm are consistent with a 50\% reduction in the collection of the scintillation and Cherenkov light. At the more extreme incident angles the Cherenkov collection efficiency is reduced by a larger factor, as the source is outside the geometric field of view of the Winston cone. The $Q_{\rm TOT}$ values are given relative to the central data, which is normalized to one.} 
    \label{tab:off-axis-results}
    \centering 
    \begin{tabular}{ccc}
         \hline\hline\noalign{\smallskip}
         y (mm) & $C^{*}_{\rm NORM}$ (1/m$^{2}$) & $Q^{\rm rel}_{\rm TOT}$ \\ 
         \noalign{\smallskip}\hline\noalign{\smallskip}
         0 & 0.91 $\pm$ 0.10  & 1.00 $\pm$ 0.06 \\ 
         75 & 0.49 $\pm$ 0.05 & 0.52 $\pm$ 0.03 \\ 
         150 & 0.06 $\pm$ 0.01 & 0.29 $\pm$ 0.02 \\ 
         \noalign{\smallskip}\hline\hline
    \end{tabular}
\end{table}

In a large detector, the geometric field of view of the dichroicon will need to be chosen given the detector design -- in particular, the size of the detector, the size and shape of the photodetectors, and the expected fiducial volume. In general, an inner fiducial volume is usually several meters away from the PMT array, thus the maximum angle of incidence on a Winston cone is quite small. The behavior of the dichroicon under far-field illumination is not measured, and will be a priority for future measurements and simulation studies.

\subsubsection{LAB+PTP and Dichroicon-2}\label{sec:results-labptp}

By using a scintillator with a shorter wavelength emission spectrum, the cut-on of the central dichroic filter can be decreased, thus increasing the total Cherenkov light collected at the aperture PMT. A fluor, PTP, was identified as being able to dissolve in LAB at 2g/L with a high light yield and a shorter wavelength emission spectrum than PPO, as presented in Figure \ref{fig:wavelengths}. Measurements discussed in this section are done with LAB+PTP and compared directly to the LAB+PPO measurements. The setup included the full dichroicon, with the R7600-U20 at the aperture, but without any absorbing longpass filter behind the dichroic filter.

The second dichroicon, called dichroicon-2, detailed in Table \ref{tab:dichroicon2} uses 462~nm longpass filter at the aperture to replace the 480~nm longpass filter. Additionally, the rectangular filters are replaced with 450~nm shortpass filters. These filters are chosen with shorter-wavelength cut-on values to reflect to and transmit through the aperture filter a larger fraction of the Cherenkov light.

\begin{figure}[t!]
\centering 
\includegraphics[width=0.45\textwidth]{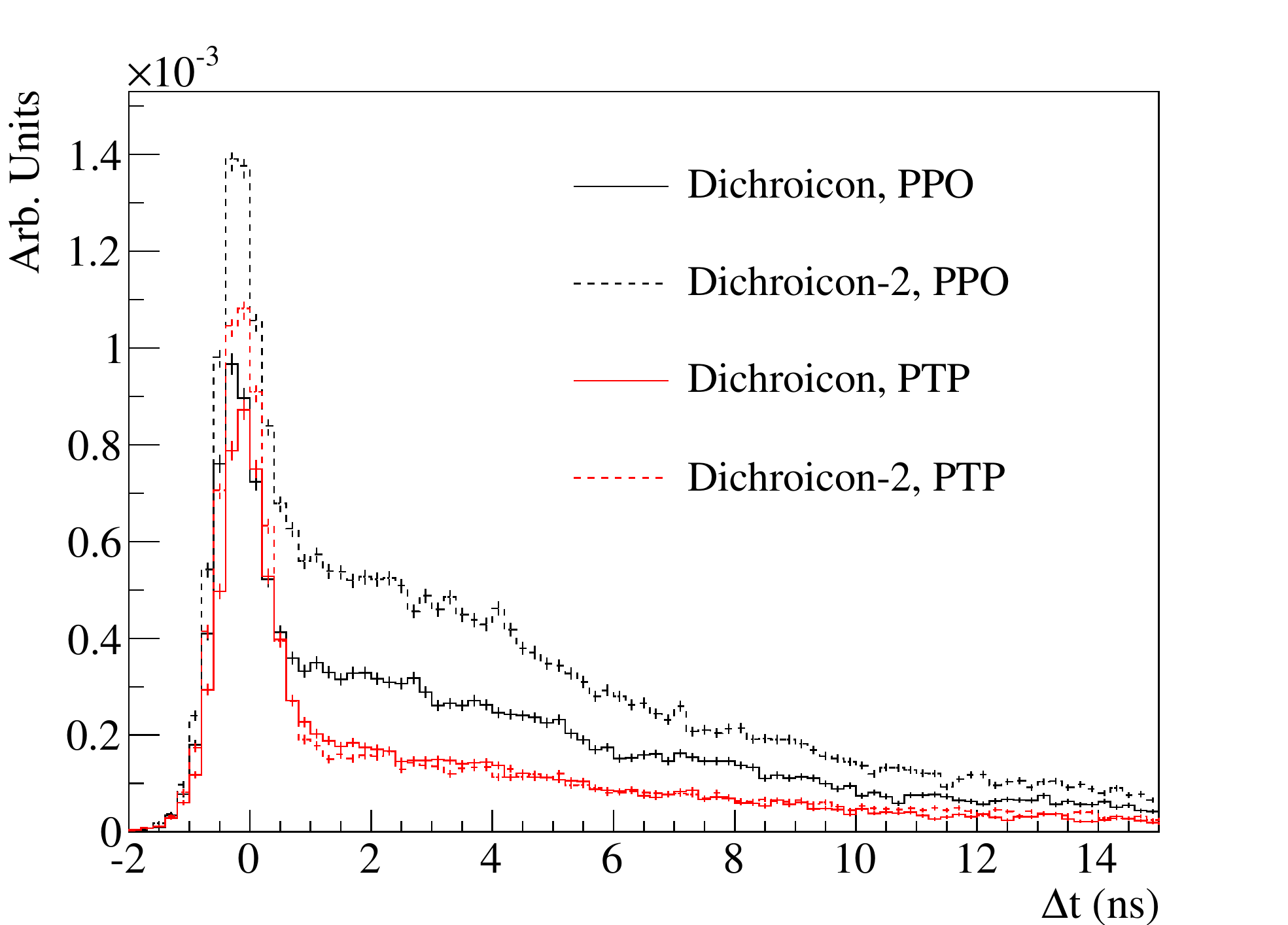}
\caption{The data for the R7600-U20 at the aperture of the dichroicon using LAB+PPO and LAB+PTP targets, compared for the dichroicon and the dichroicon-2.}
\label{fig:dichroicon2}
\end{figure}

The data for the R7600-U20 PMT with a LAB+PPO source, shown in Figure \ref{fig:dichroicon2}, shows an increase in the amount of Cherenkov and scintillation light detected when using the dichroicon-2. This is shown similarly for the LAB+PTP source in Figure \ref{fig:dichroicon2}; however, given the shorter wavelength emission spectrum of the PTP, there is only an increase in the Cherenkov light detected, with no change to the amount of scintillation light leaking through the dichroic filter.

These results are presented quantitatively in Table \ref{tab:dichroicon-comp}. The definitions of $C^{*}_{\rm NORM}$, $P$, $R$, and $Q_{\rm TOT}$ are the same as the ones presented in Section \ref{sec:results-labppo}. The first row shows the LAB+PPO results already presented in Table \ref{tab:central-fits} for the R7600-U20 PMT at the aperture of the dichroicon. By replacing the LAB+PPO with LAB+PTP the amount of Cherenkov light, quantified by $C^{*}_{\rm NORM}$, does not change, as expected, but the purity of the selection in a prompt window improves to about 88\%. This improvement comes from the smaller amount of scintillation light leakage through the central filter, due to the shorter emission spectrum, and thus less scintillation light in the selected prompt window. 

\begin{table*}[t!]
    \caption{Comparison between LAB+PPO and LAB+PTP and the two different dichroicons for the R7600-U20 aperture PMT. For both scintillator cocktails, using the dichroicon-2 with shorter wavelength pass filter increased $R$.}
    \label{tab:dichroicon-comp}
    \centering
    \begin{tabular}{cccccc}
        \hline\hline\noalign{\smallskip}
         Scintillator & Dichroicon & $C^{*}_{\rm NORM}$ (1/m$^{2}$) & $P$ & $R$ & $Q_{\rm TOT}$ (10$^{6}$ pC) \\
         \noalign{\smallskip}\hline\noalign{\smallskip}
         LAB+PPO & 1 & 8.71 $\pm$ 0.64 & 84.4 $\pm$ 1.5 & 7.35 $\pm$ 0.56 & 9.47 $\pm$ 0.56 \\  
         LAB+PTP & 1 & 8.94 $\pm$ 0.67 & 88.2 $\pm$ 1.6 & 7.89 $\pm$ 0.61 & 7.03 $\pm$ 0.30 \\
         LAB+PPO & 2 & 12.38 $\pm$ 0.93 & 81.2 $\pm$ 1.9 & 10.05 $\pm$ 0.79 & 9.03 $\pm$ 0.54 \\  
         LAB+PTP & 2 & 12.54 $\pm$ 0.92 & 92.0 $\pm$ 1.4 & 11.54 $\pm$ 0.86 & 7.12 $\pm$ 0.33 \\
        \noalign{\smallskip}\hline\hline
    \end{tabular}
\end{table*}

Rows three and four in Table \ref{tab:dichroicon-comp} show the results for dichroicon-2, which show a notable increase of about 40\% in the total Cherenkov light collected. Again, this is consistent between the LAB+PPO and LAB+PTP results. The purity of the selection for the LAB+PPO data decreases as there is more scintillation light leakage through the shorter wavelength central filter. However, there is no increase in the scintillation light leakage for the LAB+PTP data, so the purity of the selection increases, as there is more Cherenkov light but the same amount of scintillation light in the prompt window. Overall, the value of $R$ for this setup with LAB+PTP and the dichroicon-2 reaches 12.54, which is the largest value for any setup tested. 

In general the dichroicon-2 performs better than the dichroicon by collecting more Cherenkov light at the aperture with no significant decrease in the purity of the Cherenkov selection. This is in part due to the narrow TTS of the R7600-U20 which allows for excellent separation, regardless of the increase in the scintillation leakage for LAB+PPO. It is also not unexpected that the performance is improved by replacing some of the filters -- the choice of the filters for the original dichroicon was not optimized. Overall, the dichroicon-2 measurements demonstrate that small, simple changes to the dichroicon can yield performance improvements, which will be further investigated in future studies. 

The data for the R1408 PMT is compared in Table \ref{tab:dichroicon-comp}, which shows the total amount of scintillation light detected in the $Q_{\rm TOT}$ column. In both cases the changes to the dichroicon do not impact the total amount of light collected by the R1408 PMT. This is expected as only a very small fraction of the scintillation light is above 450~nm, so replacing half of the shortpass filters has very little impact on the total scintillation light transmitted through the dichroicon.

The disadvantage of using LAB+PTP can be understood by comparing the $Q_{\rm TOT}$ column between LAB+PPO and LAB+PTP. It is clear that using the LAB+PTP the total collected scintillation light at the R1408 is about 75\% of that for LAB+PPO. Part of this effect comes from the QE of the R1408, shown in Figure \ref{fig:wavelengths}. However the QE of the R1408 is fairly flat across the emission spectra of LAB+PPO and LAB+PTP and does not explain the majority of the effect. The amount of light collected at the R7600-U20 trigger PMT is a good indicator of the total light yield of the scintillator. This comparison is again done by integrating the charge spectra, and shows that the LAB+PTP light yield is about 20\% lower than the LAB+PPO light yield. This is consistent with the lower amount of collected scintillation light for the R1408 PMT, indicating that we are not losing any additional scintillation light in our setup by using LAB+PTP. Rather it is simply the intrinsic light output of the scintillator which appears to be lower. This is not unexpected because LAB+PPO is a very popular liquid scintillator specifically for having a very high light yield.

\section{Large-Scale Detector Simulation}\label{sec:large-scale-simulation}

\begin{figure}[b!]
\centering 
\includegraphics[width=\columnwidth]{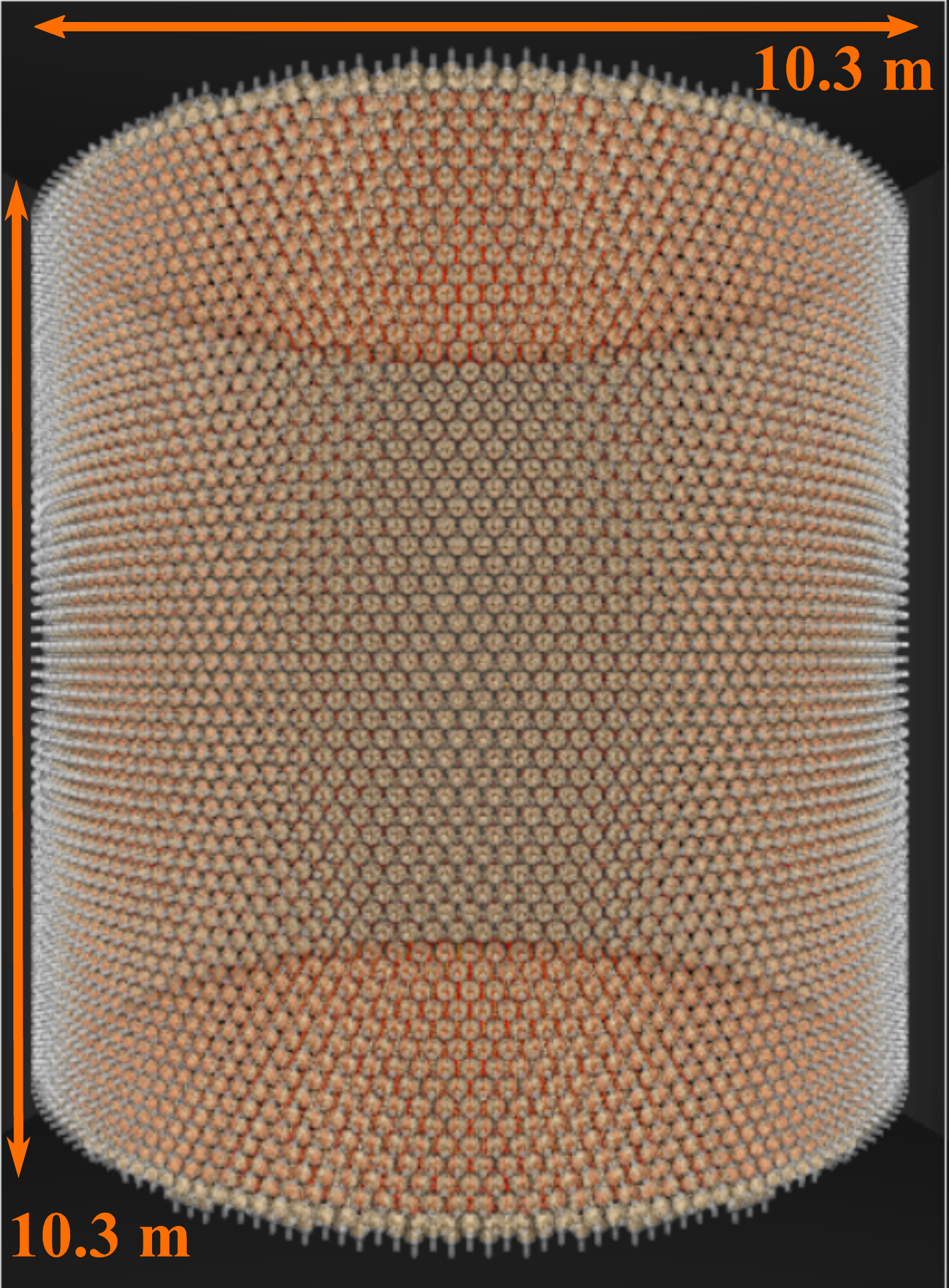}
\caption{A visualization of a 1-kT right cylinder active volume of LAB+PPO instrumented with 13,350 dichroicons produced by \texttt{Chroma}.}
\label{fig:dichroic-theia}
\end{figure}

\begin{figure}[b!]
\centering 
\includegraphics[width=\columnwidth]{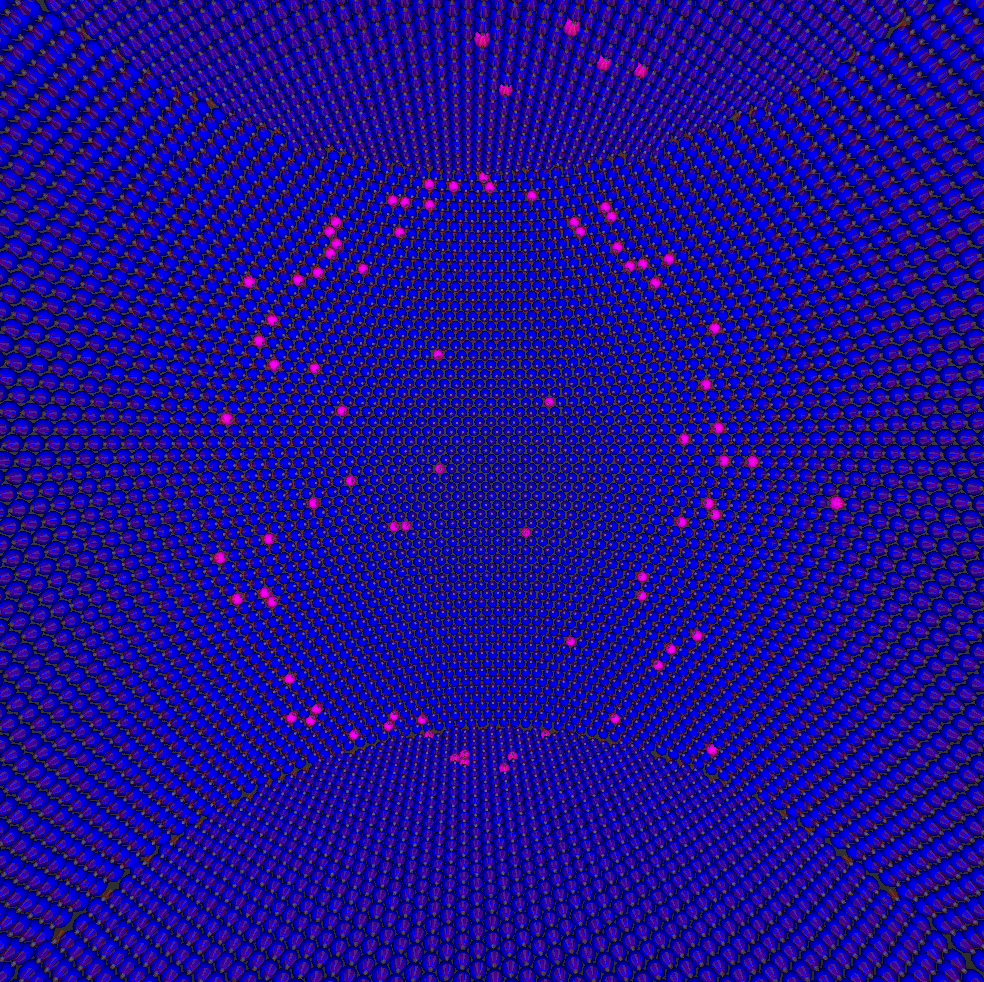}
\caption{A \texttt{Chroma} event display showing a single 100 MeV electron event in LAB+PPO. Dichroicons are colored blue if a short-wavelength hit was detected, or red if a long-wavelength hit was detected. Both a long- and short-wavelength hit results in a magenta dichroicon. Despite all dichroicons detecting many short-wavelength scintillation photons, a clear long-wavelength Cherenkov ring can be seen in magenta. Every dichroicon detects at least one short-wavelength photon so there are no dichroicons colored red in this image. No selection criterion was used when choosing this event and most events appear similar during hand-scanning of the simulated data.}
\label{fig:dichroic-theia-event}
\end{figure}

The \texttt{Chroma} model for the dichroicon can be used to simulate large-scale detectors and evaluate the efficacy of dichroicons. A basic model consisting of a 1-kT right cylinder active volume of LAB+PPO surrounded by 13,350 dichroicons is shown in Figure \ref{fig:dichroic-theia}. The dichroicon model is the same as shown in Figure \ref{fig:chroma-benchtop-model}. A single 100 MeV electron event is shown in Figure \ref{fig:dichroic-theia-event} where dichroicons are colored blue if only a short-wavelength hit was detected, red if only a long-wavelength hit is detected, and magenta if both a short and long-wavelength hit was detected. This shows a clear Cherenkov ring on the long-wavelength PMTs despite every short-wavelength PMT being hit with scintillation photons, illustrating the usefulness of the dichroicon detection scheme. Using this simulation model, the dichroicon can be optimized for maximal physics performance in large-scale detectors, which is the focus of future studies.

\section{Conclusion}

In this paper we have introduced the dichroicon, a Winston-style light concentrator built from dichroic filters, designed to sort photons by wavelength in large-scale photon-based detectors. For very large water Cherenkov detectors, the dichroicon can provide measurements of photon dispersion across large distances, and thus improve event position reconstruction and timing. As we have shown, the spectral sorting of Cherenkov photons works well (for our low-energy $^{90}$Sr source) and the model of the dichroicon response is reasonably well-understood. We have also shown that, for scintillator detectors, even in the presence of high light-yield scintillator like LAB-PPO, we can observe Cherenkov photons with purities near 90\%, while also observing a large fraction of the scintillation light.

Thus the dichroicon provides a way in principle to create truly hybrid Cherenkov/scintillation detectors, in which a very broad range of physics can be done. While there are other ways to observe both Cherenkov and scintillation light in a liquid scintillator detector, the dichroicon approach has the advantage of allowing high scintillation light yield---important for low-energy physics---while retaining the fast timing of the scintillator and Cherenkov light, and with high purity of the latter. Monolithic neutrino detectors are not the only possible application: dichroicons can also be used with segmented scintillation detectors if there is an interest in observing Cherenkov light distinct from scintillation light. In any detector in which detection area is limited---either at the front-face of a detector segment or on the walls of a monolithic detector---the dichroicon provides an effective way of sorting photons by wavelength.

We have included here also transmission and reflection measurements of the dichroic filters as a function of incidence angle and wavelength, and these measurements were used as input into \texttt{Chroma}, a fast photon ray-tracer. \texttt{Chroma} allowed for a detailed model of the dichroicon to be built and the bench-top setup simulated, the results of which how good agreement with the Cherenkov source data. Initial studies using \texttt{Chroma} to model large-scale detectors have begun and show promising initial results. 

Our measurements with liquid scintillator sources demonstrated Cherenkov and scintillation light separation with a maximum Cherenkov purity of 93.2 $\pm$ 1.3\%. Additionally, measurements made with an $\alpha$ source demonstrated the ability to perform PSD with liquid scintillators by using the absence of Cherenkov light, a possibility that helps eliminate $\beta$-$\alpha$ backgrounds from decays like $^{214}$Bi or $^{212}$Bi. With a second dichroicon, constructed with slightly different filters, we found better performance than our first prototype, and found the best Cherenkov/scintillation separation occured with this ``dichroicon-2'' and LAB-PTP, which has a narrower band of secondary emission than LAB-PPO.

Optimization of the dichroicon will depend in part on the goals of any particular experiment, as the phase space for this optimization depends on the required angular acceptance (thus detector fiducial volume), the anticipated wavelength spectrum (thus fluors if present), required detector pixelization, and overall light yield of Cherenkov and/or scintillation photons. We expect in a future publication to include a more complete \texttt{Chroma} model for large-scale detectors, based on a more integrated dichroicon design, and studies of reconstruction and particle ID for various physics goals.

\section*{Acknowledgements}

We are grateful to Gabriel Orebi Gann for helpful discussion about Cherenkov and scintillation light separation and future application with Theia, and to Jose Maneira who reminded us that Winston cones are optimal non-imaging light collectors. Thanks to Gene Beier, Ian Coulter, Eric Marzec, Andy Mastbaum, and Nuno Barros for useful discussion regarding the LAB+PPO timing and emission spectra, and Logan Lebanowski who pointed out the possibility of nesting the photon sensors. We are also grateful to Carlos Escobar and Flavio Cavanna for helpful information about dichroic filters. Thanks to Tony LaTorre for developing the Lecrunch software used for the data aquisition and for suggestions regarding fitting the timing spectrum. With respect to future use in the WATCHMAN and ANNIE detectors, we are grateful for conversations with Adam Bernstein and his group at Lawrence Livermore National Laboratory, Christopher Mauger and his group at the University of Pennsylvania, and to Bob Svoboda and his group at the University of California at Davis. Thanks to Steve Biller and Mark Chen for useful discussions regarding various scintillators and fluors. The authors would like to thank the SNO+ collaboration for providing data for the scintillator model used in our simulation software. The 3D printed object is printed courtesy of the Biomedical Library of the University of Pennsylvania. This work was supported by the Department of Energy and the Office of High Energy Physics, under grant number DE-FG02-88ER40479.

\bibliographystyle{unsrtnat}
\bibliography{bibliography.bib}

\begin{thebibliography}{67}
\providecommand{\natexlab}[1]{#1}
\providecommand{\url}[1]{\texttt{#1}}
\expandafter\ifx\csname urlstyle\endcsname\relax
  \providecommand{\doi}[1]{doi: #1}\else
  \providecommand{\doi}{doi: \begingroup \urlstyle{rm}\Url}\fi

\bibitem[Becker-Szendy et~al.(1995)]{IMB}
R.~Becker-Szendy et~al.
\newblock {Neutrino measurements with the IMB detector}.
\newblock \emph{Nucl. Phys. Proc. Suppl.}, 38:\penalty0 331--336, 1995.
\newblock \doi{10.1016/0920-5632(94)00765-N}.

\bibitem[Ahn et~al.(2006)]{Ahn:2006zza}
M.~H. Ahn et~al.
\newblock {Measurement of Neutrino Oscillation by the K2K Experiment}.
\newblock \emph{Phys. Rev.}, D74:\penalty0 072003, 2006.
\newblock \doi{10.1103/PhysRevD.74.072003}.

\bibitem[Fukuda et~al.(1998)]{Fukuda:1998mi}
Y.~Fukuda et~al.
\newblock {Evidence for oscillation of atmospheric neutrinos}.
\newblock \emph{Phys. Rev. Lett.}, 81:\penalty0 1562--1567, 1998.
\newblock \doi{10.1103/PhysRevLett.81.1562}.

\bibitem[Ahmad et~al.(2002)]{Ahmad:2002jz}
Q.~R. Ahmad et~al.
\newblock {Direct evidence for neutrino flavor transformation from neutral
  current interactions in the Sudbury Neutrino Observatory}.
\newblock \emph{Phys. Rev. Lett.}, 89:\penalty0 011301, 2002.
\newblock \doi{10.1103/PhysRevLett.89.011301}.

\bibitem[Eguchi et~al.(2003)]{kamland}
K.~Eguchi et~al.
\newblock {First results from KamLAND: Evidence for reactor anti-neutrino
  disappearance}.
\newblock \emph{Phys. Rev. Lett.}, 90:\penalty0 021802, 2003.
\newblock \doi{10.1103/PhysRevLett.90.021802}.

\bibitem[An et~al.(2012)]{An:2012eh}
F.~P. An et~al.
\newblock {Observation of electron-antineutrino disappearance at Daya Bay}.
\newblock \emph{Phys. Rev. Lett.}, 108:\penalty0 171803, 2012.
\newblock \doi{10.1103/PhysRevLett.108.171803}.

\bibitem[Ahn et~al.(2012)]{Ahn:2012nd}
J.~K. Ahn et~al.
\newblock {Observation of Reactor Electron Antineutrino Disappearance in the
  RENO Experiment}.
\newblock \emph{Phys. Rev. Lett.}, 108:\penalty0 191802, 2012.
\newblock \doi{10.1103/PhysRevLett.108.191802}.

\bibitem[Abe et~al.(2011)]{Abe:2011sj}
K.~Abe et~al.
\newblock {Indication of Electron Neutrino Appearance from an
  Accelerator-produced Off-axis Muon Neutrino Beam}.
\newblock \emph{Phys. Rev. Lett.}, 107:\penalty0 041801, 2011.
\newblock \doi{10.1103/PhysRevLett.107.041801}.

\bibitem[Adamson et~al.(2017)]{Adamson:2017gxd}
P.~Adamson et~al.
\newblock {Constraints on Oscillation Parameters from $\nu_e$ Appearance and
  $\nu_\mu$ Disappearance in NOvA}.
\newblock \emph{Phys. Rev. Lett.}, 118\penalty0 (23):\penalty0 231801, 2017.
\newblock \doi{10.1103/PhysRevLett.118.231801}.

\bibitem[Aartsen et~al.(2014)]{Aartsen:2014gkd}
M.~G. Aartsen et~al.
\newblock {Observation of High-Energy Astrophysical Neutrinos in Three Years of
  IceCube Data}.
\newblock \emph{Phys. Rev. Lett.}, 113:\penalty0 101101, 2014.
\newblock \doi{10.1103/PhysRevLett.113.101101}.

\bibitem[Aguilar-Arevalo et~al.(2013)]{Aguilar-Arevalo:2013pmq}
A.~A. Aguilar-Arevalo et~al.
\newblock {Improved Search for $\bar \nu_\mu \rightarrow \bar \nu_e$
  Oscillations in the MiniBooNE Experiment}.
\newblock \emph{Phys. Rev. Lett.}, 110:\penalty0 161801, 2013.
\newblock \doi{10.1103/PhysRevLett.110.161801}.

\bibitem[Bellini et~al.(2011)]{Bellini:2011rx}
G.~Bellini et~al.
\newblock {Precision measurement of the 7Be solar neutrino interaction rate in
  Borexino}.
\newblock \emph{Phys. Rev. Lett.}, 107:\penalty0 141302, 2011.
\newblock \doi{10.1103/PhysRevLett.107.141302}.

\bibitem[Dalmasson et~al.(2018)Dalmasson, Gratta, Jamil, Kravitz, Malek, Wells,
  Bentley, Steven, and Su]{Dalmasson:2017pow}
Jacopo Dalmasson, Giorgio Gratta, Ako Jamil, Scott Kravitz, Milad Malek, Kevin
  Wells, Julie Bentley, Samuel Steven, and Jiani Su.
\newblock {Distributed Imaging for Liquid Scintillation Detectors}.
\newblock \emph{Phys. Rev.}, D97\penalty0 (5):\penalty0 052006, 2018.
\newblock \doi{10.1103/PhysRevD.97.052006}.

\bibitem[Kaptanoglu(2018)]{Kaptanoglu:2017jxo}
Tanner Kaptanoglu.
\newblock {Characterization of the Hamamatsu 8" R5912-MOD Photomultiplier
  Tube}.
\newblock \emph{Nucl. Instrum. Meth.}, A889:\penalty0 69--77, 2018.
\newblock \doi{10.1016/j.nima.2018.01.086}.

\bibitem[Orebi~Gann(2015)]{Gann:2015fba}
Gabriel~D. Orebi~Gann.
\newblock {Physics Potential of an Advanced Scintillation Detector: Introducing
  THEIA}.
\newblock \emph{arXiv:1504.08284 [physics.ins-det]}, 2015.

\bibitem[Bonventre and Orebi~Gann(2018)]{Bonventre:2018hyd}
R.~Bonventre and G.~D. Orebi~Gann.
\newblock {Sensitivity of a low threshold directional detector to CNO-cycle
  solar neutrinos}.
\newblock \emph{Eur. Phys. J.}, C78\penalty0 (6):\penalty0 435, 2018.
\newblock \doi{10.1140/epjc/s10052-018-5925-7}.

\bibitem[Orebi~Gann(2019)]{OrebiGann:2019ncf}
G.~D. Orebi~Gann.
\newblock {Solar neutrinos with Theia}.
\newblock In \emph{{Proceedings, 5th International Solar Neutrino Conference:
  Dresden, Germany, June 11-14, 2018}}, pages 345--361, 2019.
\newblock \doi{10.1142/9789811204296_0021}.

\bibitem[Elagin et~al.(2017)Elagin, Frisch, Naranjo, Ouellet, Winslow, and
  Wongjirad]{Elagin:2016zgp}
Andrey Elagin, Henry Frisch, Brian Naranjo, Jonathan Ouellet, Lindley Winslow,
  and Taritree Wongjirad.
\newblock {Separating Double-Beta Decay Events from Solar Neutrino Interactions
  in a Kiloton-Scale Liquid Scintillator Detector By Fast Timing}.
\newblock \emph{Nucl. Instrum. Meth.}, A849:\penalty0 102--111, 2017.
\newblock \doi{10.1016/j.nima.2016.12.033}.

\bibitem[Jiang and Elagin(2019)]{Jiang:2019cnb}
Runyu Jiang and Andrey Elagin.
\newblock {Space-Time Discriminant to Separate Double-Beta Decay from $^8$B
  Solar Neutrinos in Liquid Scintillator}.
\newblock \emph{arXiv:1902.06912 [physics.ins-det]}, 2019.

\bibitem[Biller(2013)]{Biller:2013wua}
Steven~D Biller.
\newblock {Probing Majorana neutrinos in the regime of the normal mass
  hierarchy}.
\newblock \emph{Phys. Rev.}, D87\penalty0 (7):\penalty0 071301, 2013.
\newblock \doi{10.1103/PhysRevD.87.071301}.

\bibitem[Caravaca et~al.(2017{\natexlab{a}})Caravaca, Descamps, Land, Wallig,
  Yeh, and Orebi~Gann]{Caravaca:2016ryf}
J.~Caravaca, F.~B. Descamps, B.~J. Land, J.~Wallig, M.~Yeh, and G.~D.
  Orebi~Gann.
\newblock {Experiment to demonstrate separation of Cherenkov and scintillation
  signals}.
\newblock \emph{Phys. Rev.}, C95\penalty0 (5):\penalty0 055801,
  2017{\natexlab{a}}.
\newblock \doi{10.1103/PhysRevC.95.055801}.

\bibitem[Caravaca et~al.(2017{\natexlab{b}})Caravaca, Descamps, Land, Yeh, and
  Orebi~Gann]{Caravaca:2016fjg}
J.~Caravaca, F.~B. Descamps, B.~J. Land, M.~Yeh, and G.~D. Orebi~Gann.
\newblock {Cherenkov and Scintillation Light Separation in Organic Liquid
  Scintillators}.
\newblock \emph{Eur. Phys. J.}, C77\penalty0 (12):\penalty0 811,
  2017{\natexlab{b}}.
\newblock \doi{10.1140/epjc/s10052-017-5380-x}.

\bibitem[Gruszko et~al.(2019)Gruszko, Naranjo, Daniel, Elagin, Gooding, Grant,
  Ouellet, and Winslow]{Gruszko:2018gzr}
Julieta Gruszko, Brian Naranjo, Byron Daniel, Andrey Elagin, Diana Gooding,
  Chris Grant, Jonathan Ouellet, and Lindley Winslow.
\newblock {Detecting Cherenkov light from 1 - 2 MeV electrons in linear
  alkylbenzene}.
\newblock \emph{JINST}, 14\penalty0 (02):\penalty0 P02005, 2019.
\newblock \doi{10.1088/1748-0221/14/02/P02005}.

\bibitem[Guo and Wang(2018)]{Guo:2018kcp}
Ziyi Guo and Zhe Wang.
\newblock {Slow Liquid Scintillator for Scintillation and Cherenkov Light
  Separation}.
\newblock \emph{Springer Proc. Phys.}, 213:\penalty0 173--177, 2018.
\newblock \doi{10.1007/978-981-13-1316-5_32}.

\bibitem[Li et~al.(2016)Li, Guo, Yeh, Wang, and Chen]{Li:2015phc}
Mohan Li, Ziyi Guo, Minfang Yeh, Zhe Wang, and Shaomin Chen.
\newblock {Separation of Scintillation and Cherenkov Lights in Linear Alkyl
  Benzene}.
\newblock \emph{Nucl. Instrum. Meth.}, A830:\penalty0 303--308, 2016.
\newblock \doi{10.1016/j.nima.2016.05.132}.

\bibitem[et~al.(2014)]{ratpac}
Stan~Seibert et~al.
\newblock {RAT User Guide}.
\newblock \url{https://rat.readthedocs.io/en/latest/}, 2014.
\newblock [Accessed Oct. 29, 2019].

\bibitem[Yeh et~al.(2011)Yeh, Hans, Beriguete, Rosero, Hu, Hahn, Diwan, Jaffe,
  Kettell, and Littenberg]{YEH201151}
M.~Yeh, S.~Hans, W.~Beriguete, R.~Rosero, L.~Hu, R.L. Hahn, M.V. Diwan, D.E.
  Jaffe, S.H. Kettell, and L.~Littenberg.
\newblock A new water-based liquid scintillator and potential applications.
\newblock \emph{Nuclear Instruments and Methods in Physics Research Section A:
  Accelerators, Spectrometers, Detectors and Associated Equipment},
  660\penalty0 (1):\penalty0 51 -- 56, 2011.
\newblock ISSN 0168-9002.
\newblock \doi{https://doi.org/10.1016/j.nima.2011.08.040}.

\bibitem[Beacom et~al.(2017)]{JinpingNeutrinoExperimentgroup:2016nol}
John~F. Beacom et~al.
\newblock {Physics prospects of the Jinping neutrino experiment}.
\newblock \emph{Chin. Phys.}, C41\penalty0 (2):\penalty0 023002, 2017.
\newblock \doi{10.1088/1674-1137/41/2/023002}.

\bibitem[Aberle et~al.(2014)Aberle, Elagin, Frisch, Wetstein, and
  Winslow]{Aberle:2013jba}
C.~Aberle, A.~Elagin, H.~J. Frisch, M.~Wetstein, and L.~Winslow.
\newblock {Measuring Directionality in Double-Beta Decay and Neutrino
  Interactions with Kiloton-Scale Scintillation Detectors}.
\newblock \emph{JINST}, 9:\penalty0 P06012, 2014.
\newblock \doi{10.1088/1748-0221/9/06/P06012}.

\bibitem[Kaptanoglu et~al.(2019)Kaptanoglu, Luo, and Klein]{Kaptanoglu:2018sus}
Tanner Kaptanoglu, Meng Luo, and Josh Klein.
\newblock {Cherenkov and Scintillation Light Separation Using Wavelength in LAB
  Based Liquid Scintillator}.
\newblock \emph{JINST}, 14\penalty0 (05):\penalty0 T05001, 2019.
\newblock \doi{10.1088/1748-0221/14/05/T05001}.

\bibitem[Winston and Enoch(1971)]{Winston:71}
Roland Winston and Jay~M. Enoch.
\newblock Retinal cone receptor as an ideal light collector.
\newblock \emph{J. Opt. Soc. Am.}, 61\penalty0 (8):\penalty0 1120--1121, Aug
  1971.
\newblock \doi{10.1364/JOSA.61.001120}.
\newblock URL
  \url{http://www.osapublishing.org/abstract.cfm?URI=josa-61-8-1120}.

\bibitem[Boger et~al.(2000)]{Boger:1999bb}
J.~Boger et~al.
\newblock {The Sudbury neutrino observatory}.
\newblock \emph{Nucl. Instrum. Meth.}, A449:\penalty0 172--207, 2000.
\newblock \doi{10.1016/S0168-9002(99)01469-2}.

\bibitem[Alimonti et~al.(2009)]{Alimonti:2008gc}
G.~Alimonti et~al.
\newblock {The Borexino detector at the Laboratori Nazionali del Gran Sasso}.
\newblock \emph{Nucl. Instrum. Meth.}, A600:\penalty0 568--593, 2009.
\newblock \doi{10.1016/j.nima.2008.11.076}.

\bibitem[Wurm et~al.(2010)]{Wurm:2010ad}
Michael Wurm et~al.
\newblock {Optical Scattering Lengths in Large Liquid-Scintillator Neutrino
  Detectors}.
\newblock \emph{Rev. Sci. Instrum.}, 81:\penalty0 053301, 2010.
\newblock \doi{10.1063/1.3397322}.

\bibitem[Photonics(2019{\natexlab{a}})]{r7600u20}
Hamamatsu Photonics.
\newblock {Hamamatsu R7600 Datasheet}.
\newblock
  \url{https://www.hamamatsu.com/resources/pdf/etd/R7600U_TPMH1317E.pdf},
  2019{\natexlab{a}}.
\newblock [Accessed Aug. 26, 2019].

\bibitem[Photonics(2019{\natexlab{b}})]{r2257}
Hamamatsu Photonics.
\newblock {Hamamatsu R2257 Datasheet}.
\newblock
  \url{https://www.hamamatsu.com/resources/pdf/etd/R2257_TPMH1141E.pdf},
  2019{\natexlab{b}}.
\newblock [Accessed Aug. 26, 2019].

\bibitem[Biller et~al.(1999)Biller, Jelley, Thorman, Fox, and
  Ward]{Biller:1999ik}
S.~D. Biller, N.~A. Jelley, M.~D. Thorman, N.~P. Fox, and T.~H. Ward.
\newblock {Measurements of photomultiplier single photon counting efficiency
  for the Sudbury Neutrino Observatory}.
\newblock \emph{Nucl. Instrum. Meth.}, A432:\penalty0 364--373, 1999.
\newblock \doi{10.1016/S0168-9002(99)00500-8}.

\bibitem[Taniguchi et~al.(2018)Taniguchi, Du, and Lindsey]{photochemCAD}
Masahiko Taniguchi, Hai Du, and Jonathan~S. Lindsey.
\newblock Photochemcad 3: Diverse modules for photophysical calculations with
  multiple spectral databases.
\newblock \emph{Photochemistry and Photobiology}, 94\penalty0 (2):\penalty0
  277--289, 2018.
\newblock \doi{10.1111/php.12862}.

\bibitem[Optics(2019{\natexlab{a}})]{edmund}
Edmund Optics.
\newblock {Edmund Optics 500~nm Shortpass Dichroic Filter}.
\newblock
  \url{https://www.edmundoptics.com/p/500nm-252-x-356mm-dichroic-shortpass-filter/23554/},
  2019{\natexlab{a}}.
\newblock [Accessed Aug. 26, 2019].

\bibitem[Optical(2019{\natexlab{a}})]{knight}
Knight Optical.
\newblock {Knight Optical 500~nm Shortpass Dichroic Filter Datasheet}.
\newblock
  \url{https://www.knightoptical.com/_public/documents/1406727233_500fdsgraph.pdf},
  2019{\natexlab{a}}.
\newblock [Accessed Aug. 26, 2019].

\bibitem[Optical(2019{\natexlab{b}})]{knightlp}
Knight Optical.
\newblock {Knight Optical 500~nm Longpass Dichroic Filter Datasheet}.
\newblock
  \url{https://www.knightoptical.com/_public/documents/1405945453_500fdlgraph.pdf},
  2019{\natexlab{b}}.
\newblock [Accessed Aug. 26, 2019].

\bibitem[Optics(2019{\natexlab{b}})]{edmund450}
Edmund Optics.
\newblock {Edmund Optics 450~nm Shorpass Dichroic Filter}.
\newblock
  \url{https://www.edmundoptics.com/p/450nm-252-x-356mm-high-performance-fluorescence}\\\url{-dichroic-filter/3964/},
  2019{\natexlab{b}}.
\newblock [Accessed Aug. 26, 2019].

\bibitem[Optical(2019{\natexlab{c}})]{knightlp480}
Knight Optical.
\newblock {Knight Optical 480~nm Longpass Dichroic Filter Datasheet}.
\newblock
  \url{https://www.knightoptical.com/_public/documents/1405673167_480fdlgraph.pdf},
  2019{\natexlab{c}}.
\newblock [Accessed Aug. 26, 2019].

\bibitem[Wetstein et~al.(2012)]{Wetstein:2012qxa}
Matthew~J. Wetstein et~al.
\newblock {Systems-Level Characterization of Microchannel Plate Detector
  Assemblies, using a Pulsed sub-Picosecond Laser}.
\newblock \emph{Phys. Procedia}, 37:\penalty0 748--756, 2012.
\newblock \doi{10.1016/j.phpro.2012.03.717}.

\bibitem[Li et~al.(2019)Li, Classen, Dazeley, Duvall, Jovanovic, Mabe, Reedy,
  and Sutanto]{LI2019162334}
Viacheslav~A. Li, Timothy~M. Classen, Steven~A. Dazeley, Mark~J. Duvall, Igor
  Jovanovic, Andrew~N. Mabe, Edward~T.E. Reedy, and Felicia Sutanto.
\newblock A prototype for sandd: A highly-segmented pulse-shape-sensitive
  plastic scintillator detector incorporating silicon photomultiplier arrays.
\newblock \emph{Nuclear Instruments and Methods in Physics Research Section A:
  Accelerators, Spectrometers, Detectors and Associated Equipment},
  942:\penalty0 162334, 2019.
\newblock ISSN 0168-9002.
\newblock \doi{https://doi.org/10.1016/j.nima.2019.162334}.

\bibitem[Rott et~al.(2017)Rott, In, Retière, and Gumplinger]{Rott:2017lip}
Carsten Rott, Seongjin In, Fabrice Retière, and Peter Gumplinger.
\newblock {Enhanced Photon Traps for Hyper-Kamiokande}.
\newblock \emph{JINST}, 12\penalty0 (11):\penalty0 P11021, 2017.
\newblock \doi{10.1088/1748-0221/12/11/P11021}.

\bibitem[Machado and Segreto(2016)]{Machado:2016jqe}
A.~A. Machado and E.~Segreto.
\newblock {ARAPUCA a new device for liquid argon scintillation light
  detection}.
\newblock \emph{JINST}, 11\penalty0 (02):\penalty0 C02004, 2016.
\newblock \doi{10.1088/1748-0221/11/02/C02004}.

\bibitem[Thorlabs(2019{\natexlab{a}})]{beamsplitter}
Thorlabs.
\newblock {50:50 Non-Polarizing Beamsplitter Cube}.
\newblock \url{https://www.thorlabs.com/thorproduct.cfm?partnumber=BS004},
  2019{\natexlab{a}}.
\newblock [Accessed Aug. 26, 2019].

\bibitem[Thorlabs(2019{\natexlab{b}})]{thorlabs}
Thorlabs.
\newblock {Unmounted LEDs}.
\newblock \url{https://www.thorlabs.com/newgrouppage9.cfm?objectgroup_id=2814},
  2019{\natexlab{b}}.
\newblock [Accessed Oct. 19, 2019].

\bibitem[LaTorre(2013)]{latorre}
Tony LaTorre.
\newblock {Lecrunch Open Source Software}.
\newblock \url{https://bitbucket.org/tlatorre/lecrunch/src/default/}, 2013.
\newblock [Accessed Nov. 19, 2018].

\bibitem[O'Keeffe et~al.(2011)O'Keeffe, O'Sullivan, and Chen]{OKeeffe:2011dex}
H.~M. O'Keeffe, E.~O'Sullivan, and M.~C. Chen.
\newblock {Scintillation decay time and pulse shape discrimination in
  oxygenated and deoxygenated solutions of linear alkylbenzene for the SNO+
  experiment}.
\newblock \emph{Nucl. Instrum. Meth.}, A640:\penalty0 119--122, 2011.
\newblock \doi{10.1016/j.nima.2011.03.027}.

\bibitem[Lombardi et~al.(2013)Lombardi, Ortica, Ranucci, and
  Romani]{Lombardi:2013nla}
Paolo Lombardi, Fausto Ortica, Gioacchino Ranucci, and Aldo Romani.
\newblock {Decay time and pulse shape discrimination of liquid scintillators
  based on novel solvents}.
\newblock \emph{Nucl. Instrum. Meth.}, A701:\penalty0 133--144, 2013.
\newblock \doi{10.1016/j.nima.2012.10.061}.

\bibitem[O'Sullivan et~al.(2012)O'Sullivan, Wan Chan~Tseung, Tolich, O'Keeffe,
  and Chen]{OSullivan:2012fwt}
E.~O'Sullivan, H.~S. Wan Chan~Tseung, N.~Tolich, H.~M. O'Keeffe, and M.~Chen.
\newblock {SNO+ Liquid Scintillator Characterization: Timing, Quenching, and
  Energy Scale}.
\newblock \emph{Nucl. Phys. Proc. Suppl.}, 229-232:\penalty0 549--549, 2012.
\newblock \doi{10.1016/j.nuclphysbps.2012.09.186}.

\bibitem[Marrodan~Undagoitia et~al.(2009)Marrodan~Undagoitia, von Feilitzsch,
  Oberauer, Potzel, Ulrich, Winter, and Wurm]{MarrodanUndagoitia:2009kq}
T.~Marrodan~Undagoitia, F.~von Feilitzsch, L.~Oberauer, W.~Potzel, A.~Ulrich,
  J.~Winter, and M.~Wurm.
\newblock {Fluorescence decay-time constants in organic liquid scintillators}.
\newblock \emph{Rev. Sci. Instrum.}, 80:\penalty0 043301, 2009.
\newblock \doi{10.1063/1.3112609}.

\bibitem[Li et~al.(2011)Li, Xiao, Cao, Li, Ruan, and Heng]{Li2011}
Xiao-Bo Li, Hua-Lin Xiao, Jun Cao, Jin Li, Xi-Chao Ruan, and Yue-Kun Heng.
\newblock Timing properties and pulse shape discrimination of {LAB}-based
  liquid scintillator.
\newblock \emph{Chinese Physics C}, 35\penalty0 (11):\penalty0 1026--1032, nov
  2011.
\newblock \doi{10.1088/1674-1137/35/11/009}.

\bibitem[DeVol et~al.(1993)DeVol, Wehe, and Knoll]{DEVOL1993354}
T.A. DeVol, D.K. Wehe, and G.F. Knoll.
\newblock Evaluation of p-terphenyl and 2,2″ dimethyl-p-terphenyl as
  wavelength shifters for barium fluoride.
\newblock \emph{Nuclear Instruments and Methods in Physics Research Section A:
  Accelerators, Spectrometers, Detectors and Associated Equipment},
  327\penalty0 (2):\penalty0 354 -- 362, 1993.
\newblock ISSN 0168-9002.
\newblock \doi{https://doi.org/10.1016/0168-9002(93)90701-I}.

\bibitem[Machado et~al.(2018)Machado, Segreto, Warner, Fauth, Gelli, Maximo,
  Pizolatti, Paulucci, and Marinho]{Machado:2018rfb}
A.~A. Machado, E.~Segreto, D.~Warner, A.~Fauth, B.~Gelli, R.~Maximo,
  A.~Pizolatti, L.~Paulucci, and F.~Marinho.
\newblock {The X-ARAPUCA: An improvement of the ARAPUCA device}.
\newblock \emph{arXiv:1804.01407 [physics.ins-det].}, 2018.
\newblock \doi{10.1088/1748-0221/13/04/C04026}.

\bibitem[Technology(2019)]{eljin}
Eljin Technology.
\newblock {Silicone Grease EJ-550}.
\newblock \url{https://eljentechnology.com/products/accessories/ej-550-ej-552},
  2019.
\newblock [Accessed Aug. 26, 2019].

\bibitem[Seiber and LaTorre(2011)]{chroma}
Stan Seiber and Tony LaTorre.
\newblock {Fast Optical Monte Carlo Simulation With Surface-Based Geometries
  Using Chroma}.
\newblock
  \url{https://pdfs.semanticscholar.org/33ad/1bae64007a43a840288a888eba7bc3e3a37a.pdf},
  2011.

\bibitem[Land(2019)]{chroma-github}
Benjamin Land.
\newblock {Chroma}.
\newblock \url{https://github.com/BenLand100/chroma}, 2019.

\bibitem[Verkerke and Kirkby(2003)]{Verkerke:2003ir}
Wouter Verkerke and David~P. Kirkby.
\newblock {The RooFit toolkit for data modeling}.
\newblock \emph{eConf}, C0303241:\penalty0 MOLT007, 2003.
\newblock [,186(2003)].

\bibitem[Yeo et~al.(2010)]{Yeo:2010zz}
I.~S. Yeo et~al.
\newblock {Measurement of the refractive index of the LAB-based liquid
  scintillator and acrylic at RENO}.
\newblock \emph{Phys. Scripta}, 82:\penalty0 065706, 2010.
\newblock \doi{10.1088/0031-8949/82/06/065706}.

\bibitem[Galbiati et~al.(2016)Galbiati, Misiaszek, and Rossi]{Galbiati:2016inv}
C.~Galbiati, M.~Misiaszek, and N.~Rossi.
\newblock {$\alpha/\beta$ discrimination in Borexino}.
\newblock \emph{Eur. Phys. J.}, A52\penalty0 (4):\penalty0 86, 2016.
\newblock \doi{10.1140/epja/i2016-16086-1}.

\bibitem[Andringa et~al.(2016)]{Andringa:2015tza}
S.~Andringa et~al.
\newblock {Current Status and Future Prospects of the SNO+ Experiment}.
\newblock \emph{Adv. High Energy Phys.}, 2016:\penalty0 6194250, 2016.
\newblock \doi{10.1155/2016/6194250}.

\bibitem[Wurm et~al.(2012)]{Wurm:2011zn}
Michael Wurm et~al.
\newblock {The next-generation liquid-scintillator neutrino observatory LENA}.
\newblock \emph{Astropart. Phys.}, 35:\penalty0 685--732, 2012.
\newblock \doi{10.1016/j.astropartphys.2012.02.011}.

\bibitem[Ashenfelter et~al.(2015)]{Ashenfelter:2015aaa}
J.~Ashenfelter et~al.
\newblock {Light Collection and Pulse-Shape Discrimination in Elongated
  Scintillator Cells for the PROSPECT Reactor Antineutrino Experiment}.
\newblock \emph{JINST}, 10\penalty0 (11):\penalty0 P11004, 2015.
\newblock \doi{10.1088/1748-0221/10/11/P11004}.

\bibitem[Amaudruz et~al.(2016)]{Amaudruz:2016qqa}
P.~A. Amaudruz et~al.
\newblock {Measurement of the scintillation time spectra and pulse-shape
  discrimination of low-energy $\beta$ and nuclear recoils in liquid argon with
  DEAP-1}.
\newblock \emph{Astropart. Phys.}, 85:\penalty0 1--23, 2016.
\newblock \doi{10.1016/j.astropartphys.2016.09.002}.

\end{thebibliography}

\end{document}